\documentclass[twocolumn,aps,epsfig,
prd,showpacs,preprintnumbers,amsmath,amssymb,amsfonts,tightenlines]{revtex4}

\usepackage{graphicx}
\usepackage{dcolumn}
\begin{document}

\preprint{ }

\def\st{\scriptstyle}
\def\sst{\scriptscriptstyle}
\def\mco{\multicolumn}
\def\epp{\epsilon^{\prime}}
\def\vep{\varepsilon}
\def\ra{\rightarrow}
\def\mee{M_{ee}}
\def\ppg{\pi^+\pi^-\gamma}
\def\pmm{\pi^+\mu^+\mu^-}
\def\kpmm{K^+\rightarrow\pi^+\mu^+\mu^-}
\def\kpme{K^+\rightarrow\pi^+\mu^+e^-}
\def\kppp{K^+\rightarrow\pi^+\pi^+ \pi^-}
\def\kaye3{K^+ \rightarrow \pi^0 e^+ \nu}
\def\ke3g{K^+\rightarrow\pi^0 e^+ \nu \gamma}
\def\kmu3{K^+\rightarrow\pi^0\mu^+ \nu}
\def\kpi2{K^+\rightarrow\pi^0\pi^+}
\def\pee{\pi^+e^+e^-}
\def\pme{\pi^+\mu^+e^-}
\def\kpee{K^+\rightarrow\pi^+e^+e^-}
\def\ke4{K^+\rightarrow\pi^+\pi^-e^+\nu}
\def\kpipi{K^+\ra \pi^+\pi^0}
\def\kmunu{K^+\ra \mu^+\nu}
\def\k3pi{K^+\ra \pi^+ \pi^0 \pi^0}
\def\ktau{K^+\ra \pi^+ \pi^+ \pi^-}
\def\eeg{e^+e^-\gamma}
\def\Mnu2{M_{\nu}^{2}}
\def\vp{{\bf p}}
\def\ko{K^0}
\def\kb{\bar{K^0}}
\def\keee{K^+ \rightarrow \pi^0 e^+ \nu}
\newcommand{\co}{\; ,}
\newcommand{\po}{\; .}
\newcommand{\nn}{\nonumber \\}  
\title{
An improved upper limit on the decay $K^+ \rightarrow \pi^+\mu^+e^-$}
\title{
An improved upper limit on the decay $K^+ \rightarrow \pi^+\mu^+e^-$}
\author{Aleksey Sher$^{7}$\cite{AS}, R. Appel$^{6,3}$,
G.S. Atoyan$^4$, B. Bassalleck$^2$,
D.R. Bergman$^{6}$\cite{DB}, N. Cheung$^3$, S. Dhawan$^6$, 
\\ H. Do$^6$,
 J. Egger$^5$, S. Eilerts$^{2}$\cite{SE}, W. Herold$^5$,
V.V. Issakov$^4$, H. Kaspar$^5$, D.E. Kraus$^3$, 
D. M. Lazarus$^1$,\\ P. Lichard$^3$, J. Lowe$^2$, J. Lozano$^6$\cite{JL},
H. Ma$^1$, W. Majid$^6$\cite{WMa},S. Pislak$^{7,6}$\cite{SP}, 
 A.A. Poblaguev$^4$,\\ P. Rehak$^1$, A. Sher$^3$\cite{ASh},
J.A. Thompson$^3$\cite{JT}, P. Tru\"ol$^{7,6}$, and M.E. Zeller$^6$   \\
}       
\affiliation{
$^1$ Brookhaven National Laboratory, Upton, NY 11973, USA\\
$^2$ Department of Physics and Astronomy,
University of New Mexico, Albuquerque, NM 87131, USA\\
$^3$ Department of Physics and Astronomy, University of Pittsburgh,
Pittsburgh, PA 15260, USA \\
$^4$ Institute for Nuclear Research of Russian Academy of Sciences,
Moscow 117 312, Russia \\
$^5$ Paul Scherrer Institut, CH-5232 Villigen, Switzerland\\
$^6$ Physics Department, Yale University, New Haven, CT 06511, USA\\
$^7$ Physik-Institut, Universit\"at Z\"urich, CH-8057 Z\"urich,
Switzerland}
\date{\today}

\begin{abstract}
Based on results of a search for the lepton-family-number-violating decay
$K^+ \rightarrow \pi^+\mu^+ e^-$ with data collected by experiment E865 at 
the Alternating Gradient Synchrotron of Brookhaven National
Laboratory (BNL), we
 place an upper limit on the branching ratio at $2.1 \times 10^{-11}$
(90\%  C.L.). Combining the results with earlier E865 data and those of a previous
 experiment, E777, an upper limit on the branching ratio of $1.3 \times 10^{-11}$ (90\% C.L.)
 is obtained.
\end{abstract}    

\maketitle

\section{\label{sec:introduction}Introduction}

 We report on a search for the decay
$\kpme$ ($K_{\pi\mu e}$).
 This is a lepton-family-number-violating (LFNV) decay and,
 thus, is strictly forbidden in
the Standard Model (SM) with massless neutrinos.
Incorporating massive neutrinos into the SM, which is required by the growing
evidence for the lepton family transformations in the neutrino 
sector\cite{McDonald:gd}, results in a prediction of 
LFNV kaon decays at an unobservably low level\cite{Langacker:1988vz}.
 Thus an observation of a LFNV process like 
the decay $K_{\pi\mu e}$ would serve as a clear
 indication of physics beyond the SM.
Moreover, extensions of the SM such as Extended Gauge
 theories\cite{Cahn:1980kv,Shanker:1981yz}, Technicolor\cite{Farhi:1980xs}
and Supersymmetry\cite{Haber:1984rc} do allow
LFNV processes. 

               A first search for $K_{\pi\mu e}$, performed during the 1970s, 
resulted in
an upper limit on the decay's branching ratio (B) of $4.8\times10^{-9}$ (90\% 
C.L.)\cite{Diamant76}.
\begin{figure}[htb!]
\resizebox{.22\textwidth}{!}{
\includegraphics{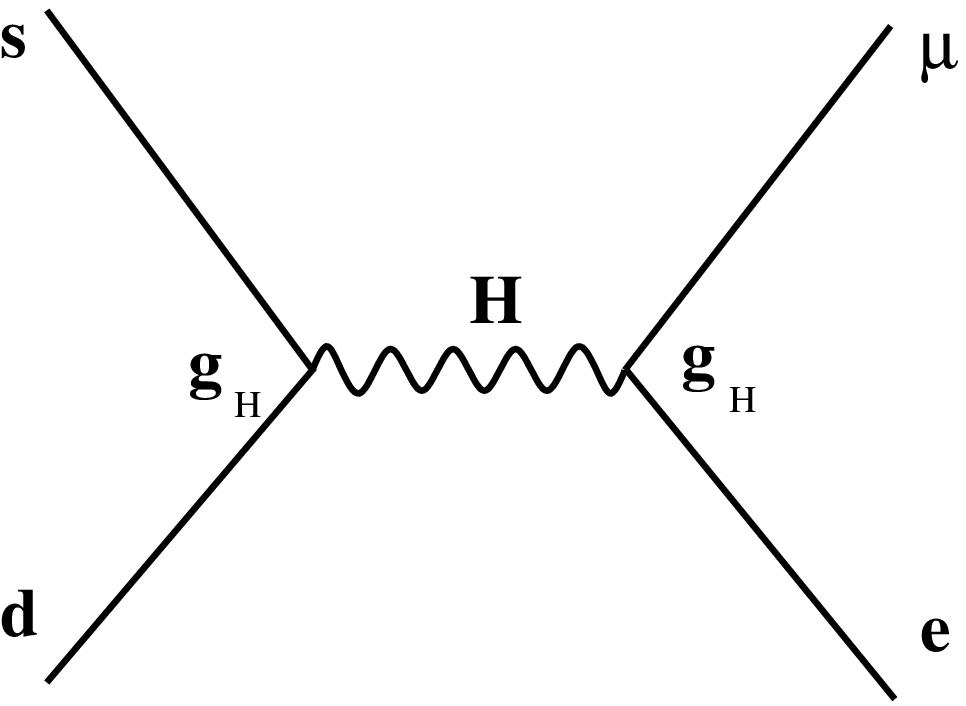}}
\hspace*{\fill}
\resizebox{.22\textwidth}{!}{
\includegraphics{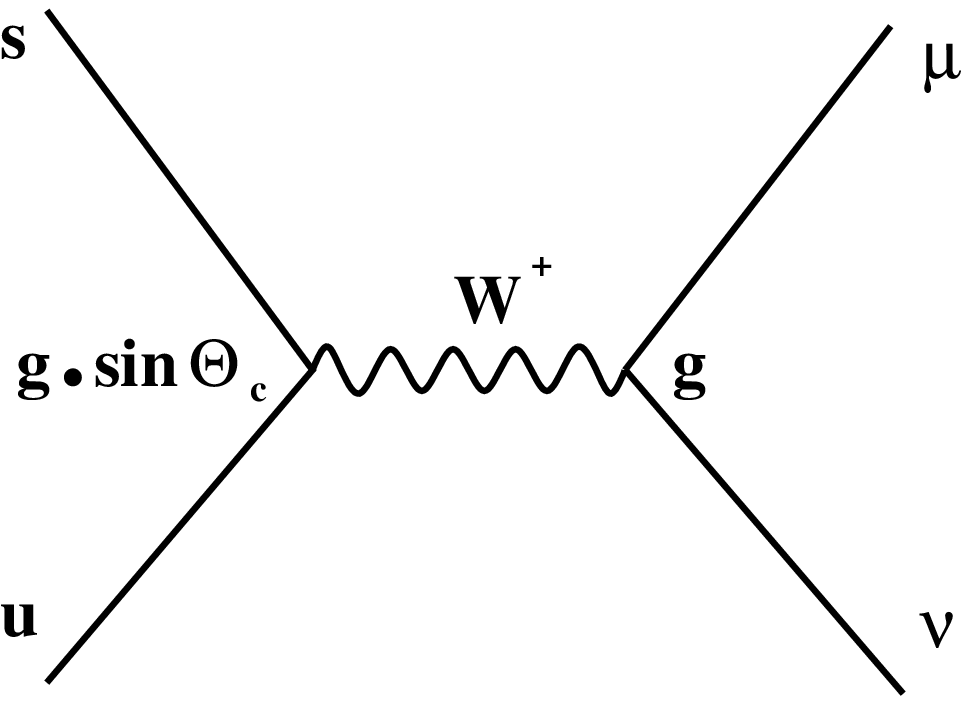}}          
\caption{ Left: Feynman diagram 
for Extended Technicolor mechanism for $\kpme$
decay; right: Feynman diagram for $\kmu3$ decay.
($\theta_{c}$ is the Cabibbo angle,
${g_{H}}$ and ${g}$ are the ETC and weak coupling constants)
   \label{fig:etc_decay}}
\end{figure}
 A
search for this decay in BNL was initiated by Experiment 777, in which the 
upper limit
$B<2.1\times10^{-10}$ was established\cite{Lee}. The data collected in
 1995\cite{sd1995}
 and 1996\cite{Hanh00,prl00}  by BNL Experiment 865, a 
successor
to E777,  allows us to lower the combined upper limit to
$B(K^+\to\pi^+\mu^+e^-)<2.8\times10^{-11}$ (90\% C.L.)\cite{prl00}.
 Nevertheless a 
null search
result is usefull since it can either put constraints on the parameters of the 
existing
extension models that allow LFNV or rule them out.

 For example, in the Extended Technicolor model (ETC)\cite{Cahn:1980kv} the
transition between the leptons of different generations can be mediated by a
horizontal ETC boson ($H$). The corresponding diagram,
 shown in Fig. 1 is similar
to the diagram of the familiar  decay $K^+\to\pi^0\mu^+\nu$. Using the
assumption that $H$ and $W$ boson couplings are approximately equal to each
other, one can relate the $H$ boson mass, $M_H$, to the  $K_{\pi\mu e}$
branching ratio\cite{Cahn:1980kv}:
\begin{equation}
    M_H \approx 85\ \mathrm{TeV}\left[\frac{10^{-11}}{B(K^+\rightarrow 
\pi^+\mu^+
e)}\right]^{1/4}
\end{equation}
The quoted limit $B(K^+\rightarrow 
\pi^+\mu^+ e)<2.8\times 10^{-11}$
corresponds to a 65 TeV lower limit on the $M_H$ mass scale. A competitive
constraint on the new physics mass scale may be obtained from another
LFNV kaon decay, $K_L\to\mu^{\pm}e^{\mp}$ ($K_{\mu e}$), for which an upper limit on its
branching ratio is set at $4.7\times10^{-12}$\cite{E791_98}.
 However, the fact that the decay
$K_{\mu e}$ is sensitive to the axial-vector  and pseudoscalar ($sd$)
quark transitions,
 while $K_{\pi\mu e}$ is sensitive to vector, scalar, and tensor
transitions, makes these two processes complementary in the search for new
phenomena.

\begin{figure*}[htb!]
\rotatebox{90}{
\resizebox{0.44\linewidth}{!}{
\includegraphics{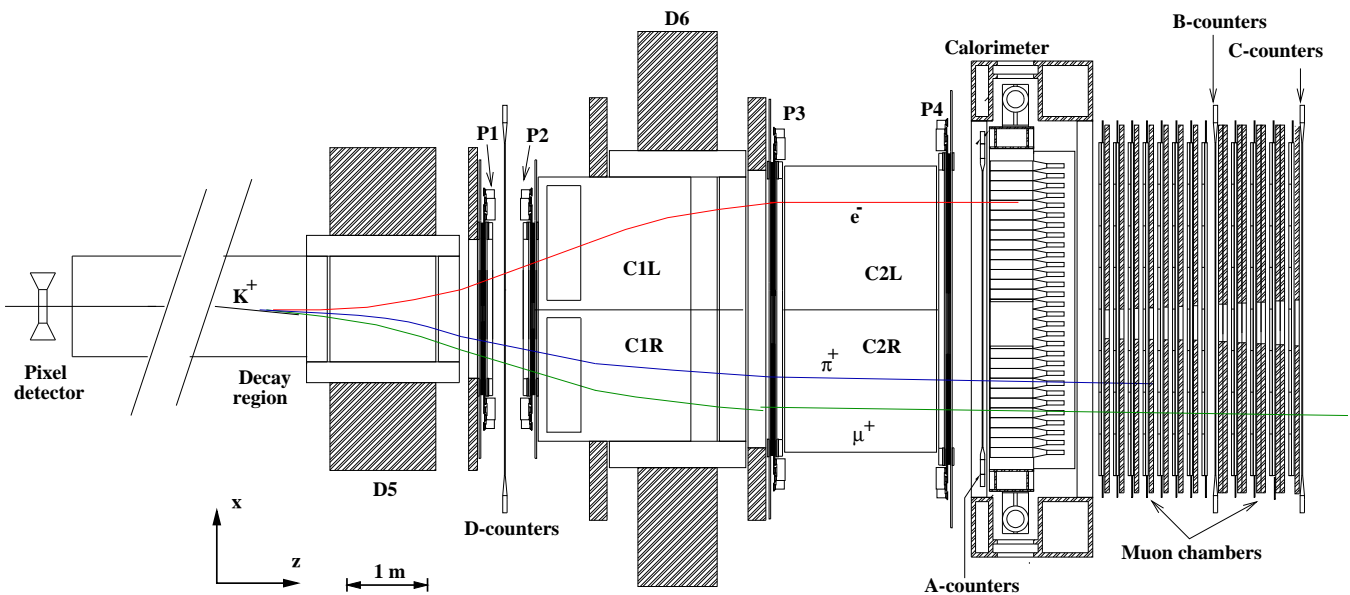}}}
\caption
{\label{detector} Overall view of the detector with a simulated $K_{\pi\mu e}$
event
(horizontal cross section at beam height). C1,C2: gas \v{C}erenkov counters;
P1,P2,P3,P4: proportional chambers; D5,D6: dipole magnets; A,B,C,D -
 scintillation counter trigger hodoscopes.}
\end{figure*}  

This illustrates the significance of the rare decay searches
which 
can be performed at an easily attainable low energy 
but at the same time can effectively 
probe the high mass region. 

The analysis outlined here is based on data recorded at the
Brookhaven Alternating Gradient Synchrotron (AGS)
during a five month run in 1998, employing the
E865 detector.
               
\section{Experimental Setup}

\subsection{Overview and design philosophy}

To perform  a successful search for a rare decay, 
like the $K_{\pi\mu e}$, two basic conditions must
be satisfied: first, the
 parent particle ($K^+$)
must be produced in copious amounts and second, the detector
must be able both to detect signal and
to suppress backgrounds to
a level low enough for a signal to be seen. The first condition was
met by employing the high intensity proton source (AGS) to
create the $K^+$ beam. 
With the projected $K_{\pi\mu e}$ single event sensitivity of
$10^{-12}$, in order
to meet the second condition we designed the E865 detector to 
have excellent event reconstruction and particle identification (PID) capabilities.
The particular requirements for the detector's event reconstruction
 and PID performance 
were derived from the careful consideration of the possible backgrounds.
  The dominant way to get a $\pi^+$, $\mu^+$ and $e^-$ from a
kaon decay is through the
decay chain $\kppp$ ($K_{\tau}$), $\pi^+ \ra \mu^+\nu$,
 $\pi^- \ra e^-\overline{\nu}$, which has a combined branching ratio of 
$6.8\times 10^{-6}$. 
 The fact that the in-flight pion decays have to happen
within the detector and the neutrinos carry off momentum,
distorting an event's kinematics,
 led to the design of a high-resolution momentum analyzing 
spectrometer which allowed
 charged particle reconstruction and momentum measurement.
The study of the Monte Carlo simulated events and
off-line event reconstruction (to be discussed later) showed that such
backgrounds could be suppressed to the level of $10^{-13}$ by applying
simple kinematic cuts.
However, the $K_{\tau}$ decay would also mimic the $K_{\pi\mu e}$ if
 a $\pi^-$ was misidentified as $e^-$ and $\pi^+$ was either 
misidentified as $\mu^+$ or
underwent an in-flight decay ($\pi^+ \ra \mu^+\nu$).
Another background was caused by
 rather common kaon decays: $\kpi2$ ($K_{\pi 2}$),
 with a $\pi^+$ misidentified as $\mu^+$, 
and  $\kmu3$ ($K_{\mu 3}$).
 Both decays would mimic $K_{\pi\mu e}$ if the $\pi^0$ underwent
a Dalitz decay ($\pi^0 \ra e^+e^-\gamma$) and the $e^+$ was misidentified as
a $\pi^+$. In the following 
we will refer to both decays ($K_{\pi 2}$ and $K_{\mu 3}$ followed by
a Dalitz decay of $\pi^0$) as $K_{Dal}$.
The suppression of the $K_{\tau}$ and $K_{Dal}$ backgrounds
 required a PID system
 with abilities to veto $\pi^-$ (while still being efficient
 in identifying $e^-$),
 veto $e^+$ and to discriminate $\mu^+$ from $\pi^+$. 

A detailed description of the apparatus (see Fig. \ref{detector})
 and the beam arrangement is presented elsewhere\cite{nimpaper}. 
Here we give only a brief overview
 of the kaon beam and detector elements.

\subsection{Kaon beam}

The AGS at Brookhaven National Laboratory served as a source of the primary 25.5 GeV/c 
proton beam, and delivered about 1$\times$ $10^{13}$  protons to the kaon production 
target in a spill of 2.8 s
 with a cycle time of 5.1 s. 
Before the extraction from the AGS, the proton beam
 was debunched, {\em i.e.} 
particles were distributed over the circumference of the
 ring to obtain a uniform beam intensity during the spill.
The extracted beam of protons, 
 hitting a 15 cm long
water-cooled copper target  with transverse dimensions
 5 $\times$ 5 $\rm{mm}^2$, produced a large variety
 of particles, but mainly pions, protons and kaons survive.
After the production target, 
the unseparated 
secondary beam was momentum selected at 6 GeV/c ($\pm 2 \%$) and was
transported to the beginning of the E865 detector
 through the specially designed 28m long A2 beam line\cite{nimpaper}. 
 A major concern during the design of the A2 beam line was 
the muon halo, produced by kaon and pion decays.
It was greatly suppressed by tight collimation
 and two
foci followed by dipole magnets, which effectively swept momentum-degraded
 muons out of the acceptance\cite{nimpaper}.
 During data taking, the beam flux
in the A2 beam line was
estimated to be 
 2 $\times$ $10^{8}$ $K^+$, 
4 $\times$ $10^9$ $\pi^+$, and 2 $\times$ $10^9$ protons per spill.

\subsection{Detector overview}
\label{sec:detector}
The E865 detector was located at the end of the
 A2 beam line,
and its first element
 was the 5-m long evacuated decay
 region within which
about 6$\%$ of the entering kaons decayed.
 Downstream of
the decay region, a dipole magnet (D5) swept the
 charged decay products away from the beam,
with negative particles going mostly to the left, and positive to the right
side of the apparatus.
The momentum-analyzing spectrometer consisted of
proportional wire chambers (PWC), P1-P4 and
a second dipole magnet (D6), which steered the particles back into the
acceptance region of the detector elements located further downstream.
The PWCs, each consisting of four wire planes, were desensitized in the
region where the beam passed.
 This arrangement yielded a momentum resolution
of $\sigma_p \approx 0.003$ $P^2$ GeV/c, where $P$, the momentum of the decay
products in GeV/c, was typically in the range 0.6 to 3.5 GeV/c.

The first part of the PID system 
consisted of two large atmospheric-pressure
 \v{C}erenkov counters (C1 and C2), located upstream and downstream of
the P3 wire chamber. Each counter was separated by a thin membrane 
into two (left
and right) parts providing independent particle identification for positive
and negative tracks and allowing use of different gases. 
 For the purpose of reducing the 
possibility of misidentifying a $\pi^-$ as  $e^-$
 the left sides (C1L, C2L) were filled
 with a high \v{C}erenkov threshold gas, hydrogen ($\gamma_t=60$),
 and had a light yield of about 2.3 photoelectrons (p.e.) for $e^-$.
To effectively register and veto $e^+$, the right sides (C1R, C2R) were filled
with a low-threshold gas, methane ($\gamma_t=35$),
 and had a light yield of about 5.8 p.e. for $e^+$.
In order to  reduce beam \v{C}erenkov radiation  in C1R and C2R,
 closed tubes filled with hydrogen gas were
placed in the beam region.

The next PID detector 
element was an electromagnetic calorimeter of the Shashlyk 
design\cite{Atoyan92}. It incorporated 582 modules, 11.4cm by 11.4cm by 15 radiation
lengths each, assembled
 in a 30$\times$20 array with 
18 modules in the middle removed for beam passage. 
The approximate energy resolution for electrons was  
8$\%$ /$\sqrt{E(\rm{GeV})}$. Typical energy deposition of
 a minimum ionizing particle was 250 MeV.

 The last PID detector element was
 a muon detection
system, that was located downstream of the calorimeter and consisted of
 24 planes of proportional tubes inter-spaced by iron plates. The 
 plate thickness was 5 cm between the first eight pairs of planes
and 10 cm between the last four.
Finally, four arrays of scintillator counter hodoscopes, A,B,C and D, 
were used for triggering purposes.

\subsection{Trigger requirements}
\label{sec:trigger}
The hardware trigger, described in detail in\cite{nimpaper}, was designed
 as a four-level structure
with increasing sophistication and response time at each level.
The lowest trigger level (T0) selected events with three 
charged-particle tracks, one on the left and two on the right side, 
by requiring at least three coincidences between an
A-counter slat and the calorimeter module behind it (A$\cdot$SH).
By using a programmable matrix lookup unit (MLU), 
for each combination of coincidences between individual counters on the
 right, only a limited,
 kinematically acceptable, region on the left was allowed. The MLU was 
programmed to maximize the acceptance of 
 the three charged body kaon decays by using simulated $K_{\pi\mu e}$ events.
In order to reduce contamination from decays occurring 
downstream of the decay region,
at least one coincidence on both left and right sides between the D-counter
 and A$\cdot$SH was required as well.
 The T0 signal was ready in about
175 ns after the
particle hit the A-counter and had a rate of about 2.5 kHz. 

 The next trigger level (T1) used information from the \v{C}erenkov counters
 and the muon range telescope 
for purpose of PID. In the case of $K_{\pi\mu e}$ the
dedicated trigger  (MUE) demanded the presence of an 
electron
and a muon.
Consequently, the MUE trigger  required
signals corresponding to at least 0.25 p.e. 
in both \v{C}erenkov counters on the left side (C1L,C2L)
 to select $e^-$ and
a hit in the muon range telescope (B-hodoscope) to be spatially consistent with the
A$\cdot$SH hit in order to select $\mu^+$. 
The T1 signal was available about 130 ns after the T0, and had a rate
 of about 1.3 kHz.

The next trigger level (T2) was designed to
discriminate between events with
high and low $e^-e^+$ invariant mass.
Information from coincident A$\cdot$SH hits, and \v{C}erenkov counters
 was used to
select events consistent with low vertical separation of $e^+e^-$ tracks
(low invariant mass). That trigger was not utilized in the $K_{\pi\mu e}$ 
search.

The final trigger level (T3) compared the 
number of hits in each of the PWC plane
with a predetermined likelihood distribution obtained offline. This 
trigger rejected a small fraction of events
 that had either too few or too many PWC hits.
During data taking, the average rate of T3 was about 1800 per 2.8 s 
 spill or about 0.6 KHz.

In addition to the MUE trigger,
 several prescaled monitor triggers were 
recorded, e.g. a minimum bias trigger (TAU), which required only T0 signal and
was dominated by the $K_{\tau}$ events, and two triggers sensitive to the 
$K_{Dal}$ events.
The first one (EEPS) demanded the presence of an $e^-$ and $e^+$
by requiring signals (at least 0.25 p.e.) in both
\v{C}erenkov counters on the left and right sides (C1L,C2L,C1R,C2R) and
was used for PID studies and background estimates.
 The second $K_{Dal}$ sensitive trigger (CERENK)
 was 
designed  to obtain an unbiased response of the
 \v{C}erenkov counters and required only
three out of four sides of the \v{C}erenkov counters (C1L,C2L,C1R,C2R)
 to have signals above 0.25 p.e.

 During a five month data taking period in 1998 we recorded 
a total of 1.3$\times 10^9$ triggers on tape, of which 0.3
$\times 10^9$ were MUE triggers, successfully 
reconstructed in the offline
analysis.

\section{$K_{\pi\mu e}$ Event selection and OFFLINE analysis}

\subsection{Event reconstruction}
The process of kinematic reconstruction of an event
followed
the T0 hardware trigger requirements (A$\cdot$SH) and started with
the ``clump'' finding, {\it i.e.}
determining positions of tracks in the calorimeter based on their
energy depositions and correlated A-counter hits.
For each ``clump'' found in the calorimeter, a window in the 
PWC P4 was defined and space points, requiring at least three of the four
 wire planes in the chamber, were searched for. 
Likewise, the space points found in
P4 determined windows in P3 and so forth. 
A track was formed if at least three PWCs contributed with a space point.
Next, using the measured magnetic field map inside the dipole magnets, 
the momentum of each track was fit. 
For events containing at least three
 reconstructed tracks, a fitting algorithm was used to determine the
decay vertex position by minimizing
the quantity $S$, the r.m.s. of the closest approach of the
three charged tracks to the common vertex point.
For events containing more than three
 tracks the combination with the smallest $S$ was tagged as the most probable 
set of track candidates from a kaon decay. 
 Studies of fully reconstructed $K_{\tau}$ events showed that
 the mean value of
$S$ increased for vertices found further upstream (lower $z$) consistent 
with Monte Carlo simulations. The dependence could be fitted with a 
second order polynomial
(denoted $\bar{S}(z)$). The normalized quantity $S_{norm}=S/\bar{S}(z)$,
 which is 
independent of $z$, was then used to judge the vertex quality in the 
$K_{\pi\mu e}$ analysis.
As the last step, the information from the PID
 detectors was assigned to each found track.

\subsection{Event Selection}

\begin{figure}
\resizebox{.45\textwidth}{!}{
\includegraphics{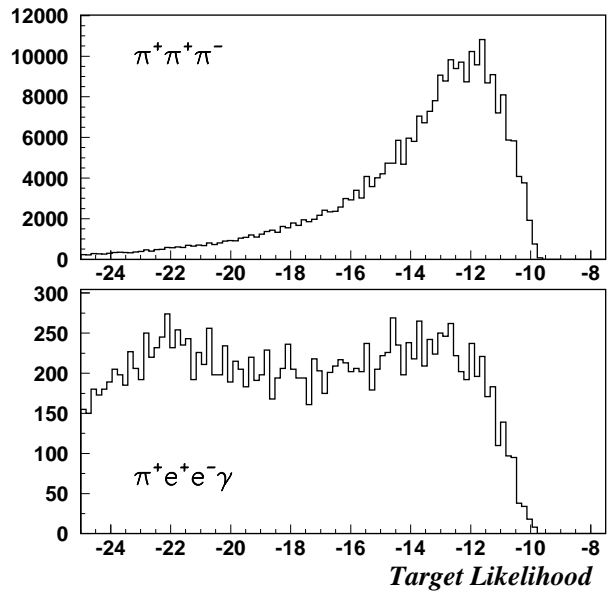}}
\caption{ Target likelihood, $L_{Target}$, distributions for
 the reconstructed $\kppp$ (upper) and $\kpi2$ with a $\pi^0 \ra e^+e^-\gamma$
 (lower) events.
 Poor target likelihood for the $\kpi2$ events is caused by
the photon, which is not included into the kinematic 
reconstruction of the kaon
 momentum
vector at the vertex causing it to fall out of
the acceptable phase space of the kaon beam. For signal events
 $L_{Target} > -20$ was required, corresponding to an
 efficiency of 92\%.
\label{fig:tgtlik}}
\end{figure}

Basic requirements for 
$K_{\pi\mu e}$ included the presence of a three-charged-track
 vertex within the decay region of an acceptable quality, $S_{norm}$.
In addition, in order to ensure that 
reconstructed tracks came from a real kaon decay with all
the particles being detected, the reconstructed total momentum vector was
tracked back to the production target using the first-order beam-line
transfer matrix, and its phase space (the kaon momentum $P_{K^+}$, 
the position $x$, $y$,
and the direction,  $\theta_x$, $\theta_y$)
 was compared to that of the kaon beam.
The five-dimensional phase space of the beam
was broken up 
into three two dimensional distributions:
 $x$ versus $\theta_x$, $y$ 
versus $\theta_y$ and
$P_{K^+}$ vs $x$ .
The properties of the kaon beam were determined from fully reconstructed
$K_{\tau}$ events and were used as input for a corresponding logarithmic 
likelihood function $L_{target}$
 (see Figures \ref{fig:tgtlik} and \ref{final_dist}, and also Sec.
 \ref{sec:likelihood}).
 The invariant mass $M_{\pi\mu e}$
, calculated using the daughter particles masses ($m_{\pi},$ $m_{\mu}$, $m_e$)
 and their measured momenta, was required to be consistent with the kaon
 mass, 493.67 MeV, within the 
 resolution, $\sigma_{M_{\pi\mu e}} \approx 4$ MeV for 
simulated $K_{\pi\mu e}$ events.

Finally we required an unambiguous identification of $e^-$, $\pi^+$ and
$\mu^+$.
Electron PID required a signal (at least 0.3 p.e.) in
the left sides of both \v{C}erenkov counters (C1L,C2L) with corrected timing
within $\pm$5 ns. In addition, the energy loss in the calorimeter ($E$) was
required to be consistent
with the momentum of the track ($P$), $i.e.$ the ratio $E/P$ to be
 between 0.65 and 1.5.
Pions were identified requiring the absence of a signal above 1.2 
p.e. on the right sides of the \v{C}erenkov counters (C2R,C2R) and
corresponding corrected timing of more than $\pm$5 ns.
In addition, energy loss in the calorimeter was required to be
consistent with that of
a minimum ionizing particle or a hadron shower ($E/P < 0.9$).
Muons were
 identified by requiring the absence of a signal above 1.2 p.e. in both
\v{C}erenkov counters (C1,C2), energy deposited in the calorimeter consistent
with minimum ionization ($E<$450 MeV) and penetration depth
 in the muon range consistent
with its momentum. Table \ref{tab:pid_pass3_eff} summarizes the
 PID efficiencies and probabilities of misidentification achieved by the above
mentioned selection cuts\cite{sher04}.

Although the selection criteria described above provided
 a high sensitivity for the signal 
(respective efficiency of each cut was approximately 95\%)
they let a considerable number of background events pass.

\begin{table}
\caption[Pass3 PID total efficiencies]
         {Identification efficiencies and probability of misidentification
          for the primary selection PID cuts.
 The symbol ``$\rightarrow$'' stands for identified as.}
\begin{ruledtabular}
\begin{tabular}{l|ccc}                                         
\hspace{.005in} & $\rightarrow \pi^+$\hspace{.5in} &
                $\rightarrow \mu^+$\hspace{.5in} &
                $\rightarrow e^-$\hspace{.5in}                \\ \hline \hline
                $\pi^+$                          &
                $0.804 \pm 0.008$                  &
                $0.066 \pm 0.001$                  &
                 -                                              \\[2mm]
                $\pi^-$                            &
                                  -                &
                -                                  &
                $(8.7 \pm 2.6) \times 10^{-6}$                       \\[3mm] 
                $\mu^+$                          &
                 -                 &
                $0.79 \pm 0.01$                  &
                 -                                              \\[3mm] 
                $e^+$                            &
                $(1.1 \pm 0.1) \times 10^{-5}$       &
                $< (1.1) \times 10^{-5}$     &
                 -                                \\[2mm] 
                 $e^-$                                  &
                 -                                 &
                 -                                 &          
                $0.767 \pm 0.003$                               \\ 
\end{tabular}
\end{ruledtabular}
\label{tab:pid_pass3_eff}
\end{table}

\subsection{Overview of backgrounds}
Possible $K_{\pi\mu e}$ backgrounds could be classified into
two different types.
 The first type consisted of rather common kaon decays and 
 was dominated by $K_{\tau}$ and $K_{Dal}$. 
 These kaon decays could be tagged as 
 $K_{\pi\mu e}$ , if kinematic requirements were satisfied and 
certain daughter particles were either misidentified or underwent
an in-flight decay as given in Table \ref{tau_dal_bg}.
Due to the excellent particle identification 
capability of the E865 apparatus, 
this type of  background could be efficiently suppressed by optimizing
the PID selection in order to 
reduce respective misidentification probabilities,
$P(\pi^-$ as $e^-)$ and $P(e^+$ as $\pi^+$).
The second type of background
 consisted of accidental
 combinations of the $\pi^+$, $\mu^+$ and $e^-$
tracks, originating from separate kaon decays,
that would satisfy kinematic
requirements for $K_{\pi\mu e}$. Such accidental events
 were usually characterized
by uncorrelated track timing and poor kinematic quality
 (vertex quality $S_{norm}$ and $L_{Target}$).
Instead 
of tightening the selection cuts to suppress
these backgrounds, we chose to employ
the likelihood method, a brief overview of which is given below.

\subsection{Overview of the Likelihood method}
\label{sec:likelihood}
In an analysis using cuts, all events that pass the final cuts
 are considered equally probable to be a signal. The
likelihood method, however, allows one to test each event against a
particular hypothesis, e.g. the likeliness for a particle to be of a 
certain type or the decay to be of a particular mode.
For example a pion, passing all the electron PID cuts, would
be misidentified as an electron.
However, for pions, the values for each of the
PID variable will most likely fall at the edge of the various PID
distributions while for a real electron they will be in the bulk.
Comparing the 
probabilities of an electron and a pion
 having  a particular PID response allows 
one to distinguish further between a real
and misidentified particle. The same method can be applied to kinematic
variables.
Thus, the use of multi-variable likelihood function allows one to
differentiate quantitatively between signal and background
 on an event-by-event basis.
\begin{table}
\caption
{\label{tau_dal_bg} Main $\kpme$ backgrounds from other $K^+$ decays.}

\begin{ruledtabular}
\begin{tabular}{ll}
$K^+$ Decay Mode              &       Misidentification \\
\hline
$\kppp$ &    \\
with $\pi^+ \ra \mu^+ \nu$ , $\pi^- \ra e^- \overline{\nu}$    & None \\[2mm]
$\kppp$  &      $\pi^-$ as $e^-$  \\
with  $\pi^+ \ra \mu^+ \nu$ &                          \\[2mm]
$\kppp$         &       $\pi^+$ as $\mu^+$   \\
with $\pi^- \ra e^- \overline{\nu}$       &                         \\[2mm]
$\kppp$         &       $\pi^+$ as $\mu^+$ and $\pi^-$ as $e^-$  \\[2mm]
$K^+ \ra \pi^+\pi^0$, $\pi^0 \ra e^+e^-\gamma$  &       $e^+$ as $\pi^+$
and $\pi^+$ as $\mu^+$ \\[2mm]
$K^+ \ra \pi^+\pi^0$, $\pi^0 \ra e^+e^-\gamma$  &       $e^+$ as $\pi^+$\\
with $\pi^+\ra\mu^+\nu$ &    \\[2mm]
$K^+ \ra \pi^0\mu^+\nu$, $\pi^0 \ra e^+e^-\gamma$       &       $e^+$ as
$\pi^+$   \\
\end{tabular}
\end{ruledtabular}
\end{table}

\label{sec:lik_form}
        The conventional definition of the log likelihood function
is
\begin{equation}
\mathcal{L}(\vec{x})=\log P(\vec{x})
\label{eq:like}
\end{equation}
 where $\vec{x}=(x_{1},x_{2}...x_{n})$ is a vector of the measured
kinematic and PID response and $P(\vec{x})$ is the n-dimensional 
probability density function (PDF). In the case where $x_{1},x_{2}...x_{n}$
are independent, the probability can be simplified into:
\begin{figure*}
\begin{center}
\centerline{
\resizebox{.33\textwidth}{!}{
\includegraphics{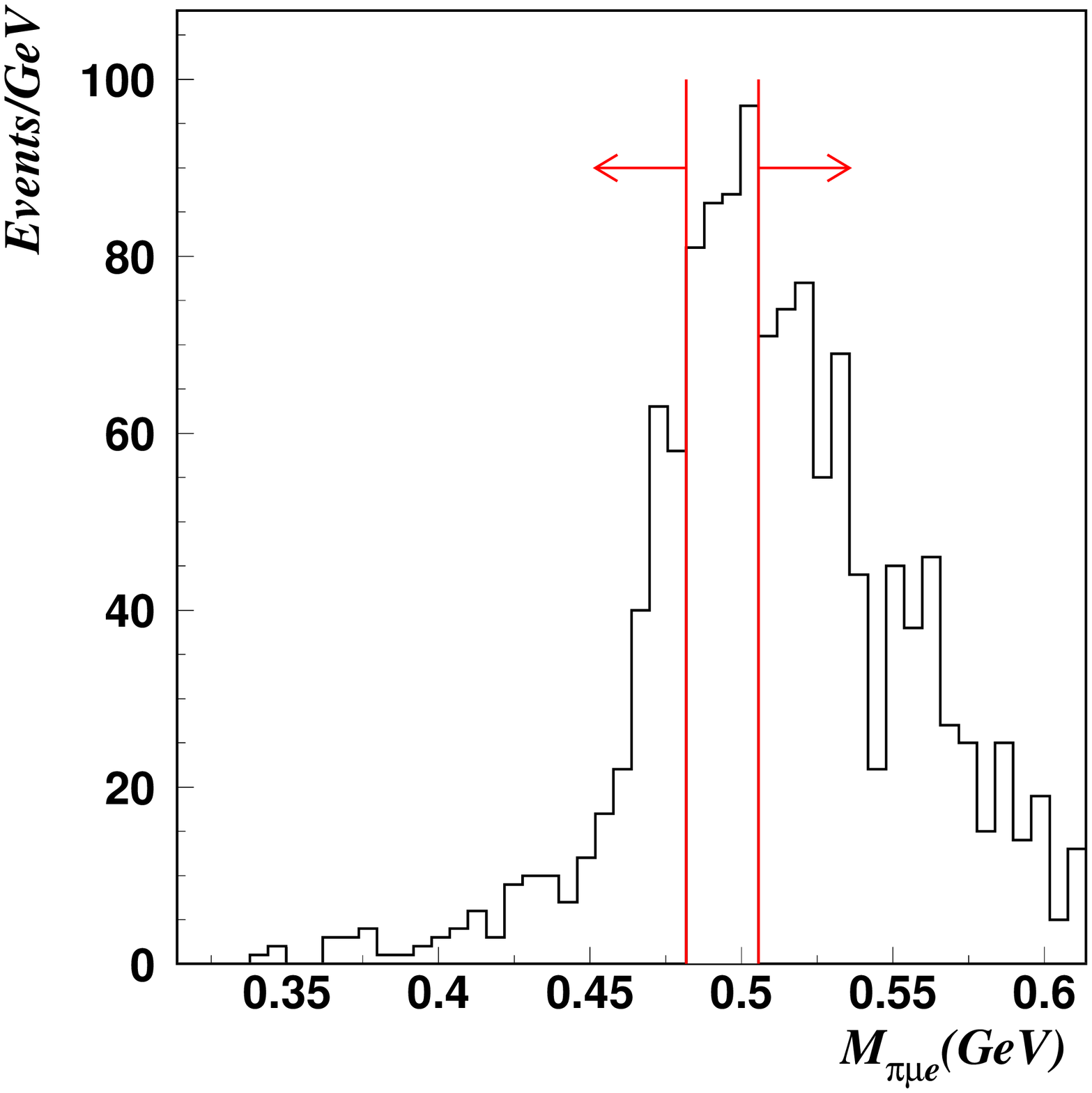}}
\hfill{}
\resizebox{.33\textwidth}{!}{
\includegraphics{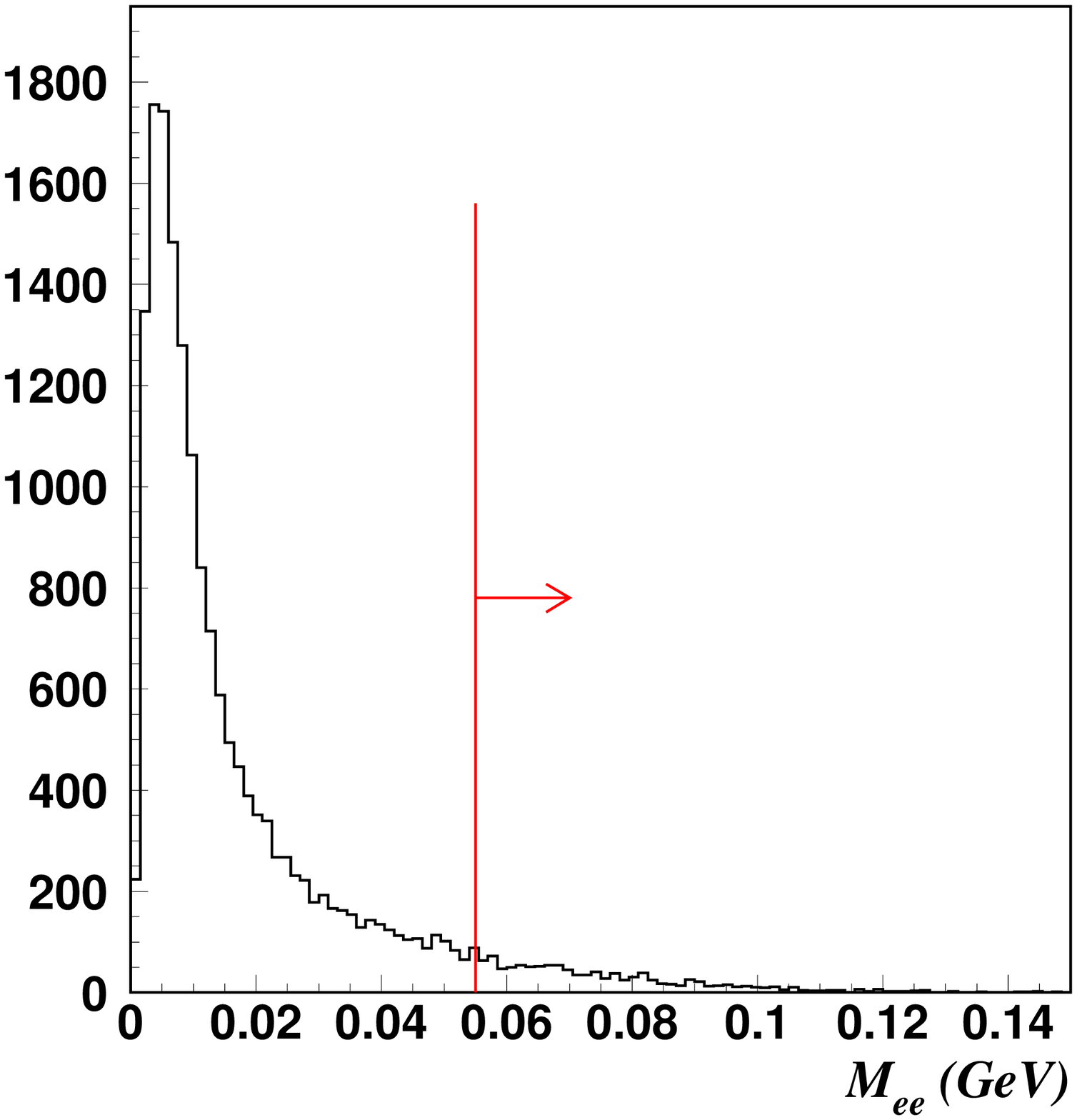}}
\hfill{}
\resizebox{.33\textwidth}{!}{
\includegraphics{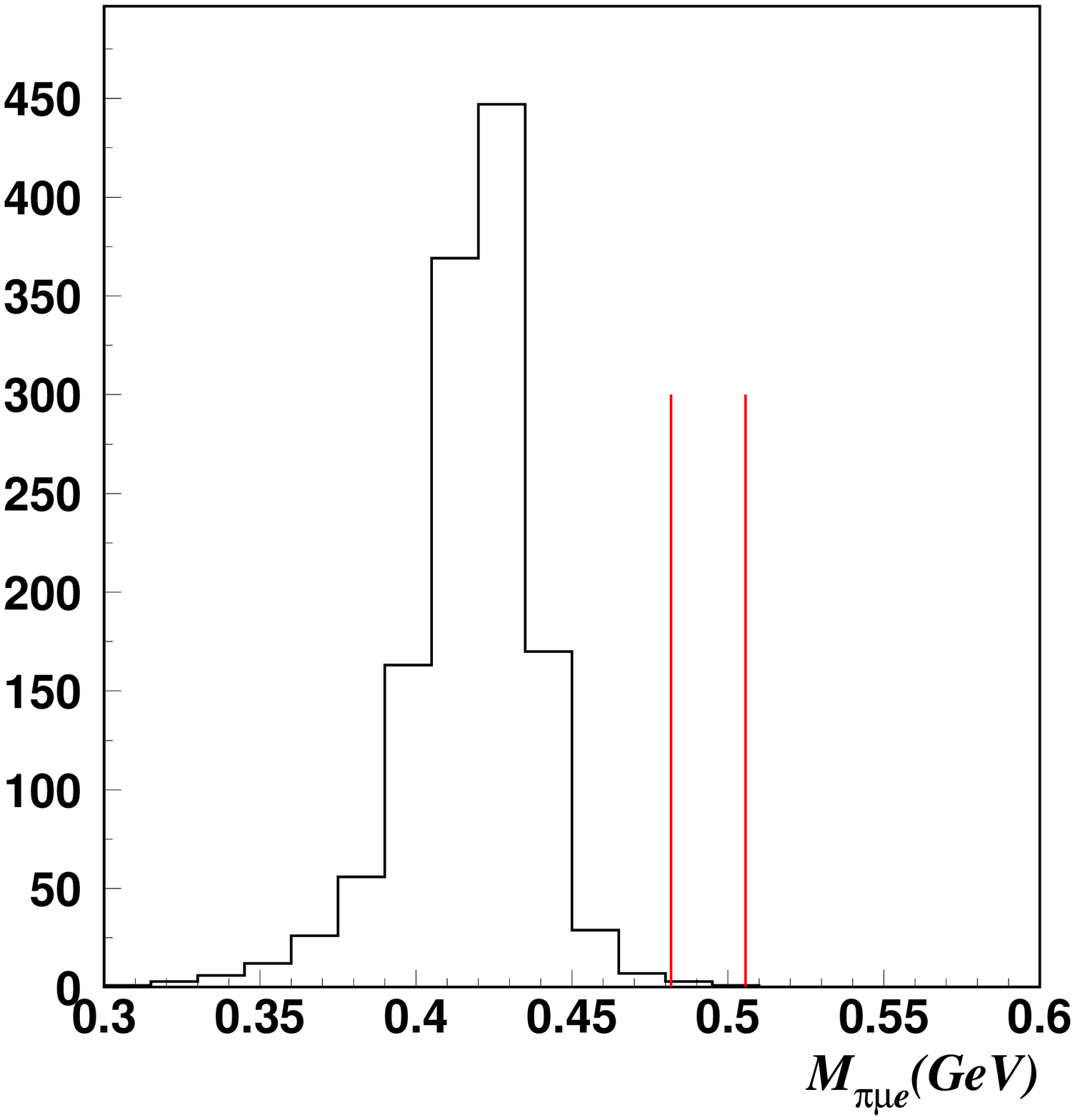}}}
\end{center}
\vspace*{-0.8cm}
\caption[Prediction of the $\kppp$ background]
{Left: $M_{\pi\mu e}$ mass distribution for $e^+\mu^+e^-$ events
measured with the EEPS monitor trigger ($K_{Dal}$ background).
Vertical lines mark the $M_{\pi\mu e}$ mass signal region;
center: $M_{ee}$ invariant mass spectrum for measured $K_ {Dal}$ events.
 The cut on $M_{ee}$ $>$ 0.055 GeV removes 94\% of the $K_{Dal}$ events;
right: $M_{\pi\mu e}$ mass spectrum for the $\pi^+\mu^+\pi^-$
 events measured with the TAU monitor trigger ($K_{\tau}$ background).
 Vertical lines mark the $M_{\pi\mu e}$ mass signal region.
}
\label{taubg_est}
\end{figure*}
\begin{equation}
\label{eq:lik_fact}
P(x_1,x_2,...,x_n) \cong P_1(x_1)\cdot P_2(x_2)\cdot...\cdot P_n(x_n)
\end{equation}
 where $P_{i}(x_{i})$ is the $i_{th}$ one dimensional
PDF,
constructed from the corresponding distribution and
$x_{i}$ is the respective measured response.
Using such an 
approximation, one can rewrite the log likelihood in the following
way:
\begin{equation}
\mathcal{L}(\vec{x})=\sum_{i=1}^{n}\log P_{i}(x_{i})
\label{eq:like1}
\end{equation}

Generally, an individual PDF can be represented as a continuous
analytical function or a binned histogram.
While it would be ideal to use the full range of a variable
 for the likelihood
function to maximize the acceptance, it was more practical
to use a finite (that would include around 95\% of signal events) range,
 choosing particularly the most
sensitive part of the distribution.
In most of the cases the range of the likelihood function was
 defined by the
values of the final cuts.
The number of bins was selected to
 be large enough to reflect the shape of
a particular distribution.
Finally, in constructing a likelihood function,
we required each PDF to be
normalized to unity.

\section{background studies}

\subsection{\boldmath $K_{Dal}$ background}

 A large part of the $K_{Dal}$ background
was rejected by applying the invariant-mass cut ($M_{ee}>55$ MeV), since
the invariant mass of the $e^+e^-$ pair from the $K_{Dal}$ decays is peaked at
low values 
(see Fig. \ref{taubg_est}).
 Events containing photons registered in the calorimeter
with a corrected timing within $\pm$ 2 ns, were rejected.
Additional rejection was
achieved by the Target Likelihood cut
 since, due to the presence of the photon that was not included
in the kinematic reconstruction,
 the total kaon momentum vector was usually inconsistent with
 that of the kaon beam.
To estimate the level of the
$K_{Dal}$ background we used the EEPS triggered data.
We selected $e^+\mu^+e^-$ events that could mimic $K_{\pi\mu e}$ (if
an $e^+$ was misidentified as a $\pi^+$)
 by applying the $K_{\pi\mu e}$
kinematic selection cuts and requiring identification of the $\mu^+$ and
$e^-$.
Events, that were selected that way, were
scaled 
by the EEPS trigger hardware prescale factor ($20$) and a 
corresponding misidentification probability - P($e^+$ as $\pi^+$). 
To reduce the misidentification probability we constructed the log-likelihood
function for the purpose of $\pi^+$ identification. 
 The likelihood included the ratio E/P and the quantity $E_{ratio}$
 describing 
the spatial spread of the electromagnetic shower in the calorimeter. The 
latter variable is defined as the ratio of the energy deposited in the 
central module, i.e. the module to which a track points, to the total 
energy, which also includes the three most adjacent modules\cite{sher04}.

The selected cut value for the $\pi^+$ PID likelihood, which resulted
 in the estimated misidentification
 probability P($e^+$ as $\pi^+$)=$(2.5\pm 0.3) \times 10^{-6}$,
ensured a satisfactory suppression
of the estimated $K_{Dal}$ background to the level of 0.09$\pm$0.01 events.

\subsection{\boldmath $K_{\tau}$ background}
\begin{figure*}
\begin{center}
\centerline{
\resizebox{.25\textwidth}{!}{
\includegraphics{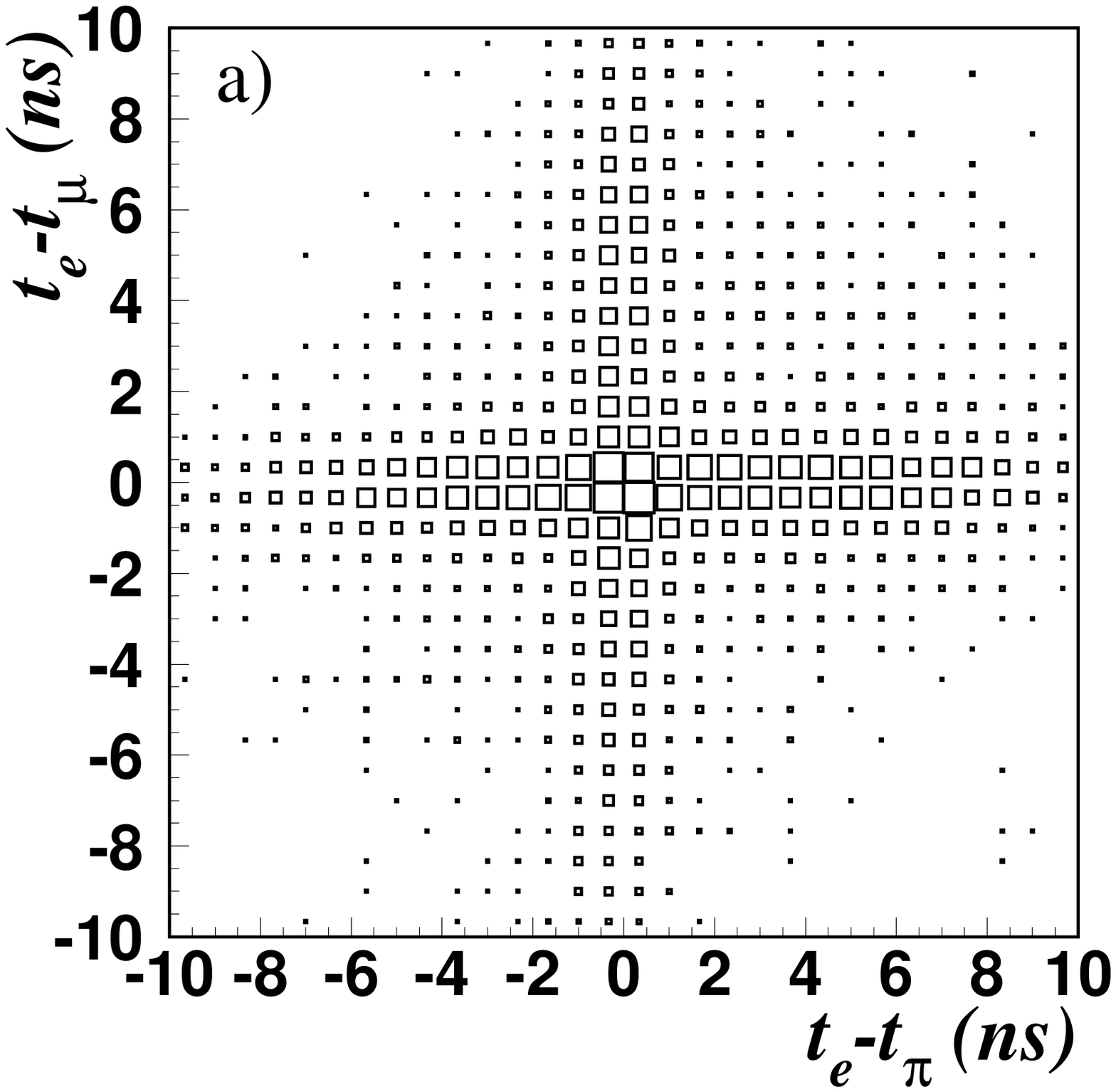}}
\resizebox{.25\textwidth}{!}{
\includegraphics{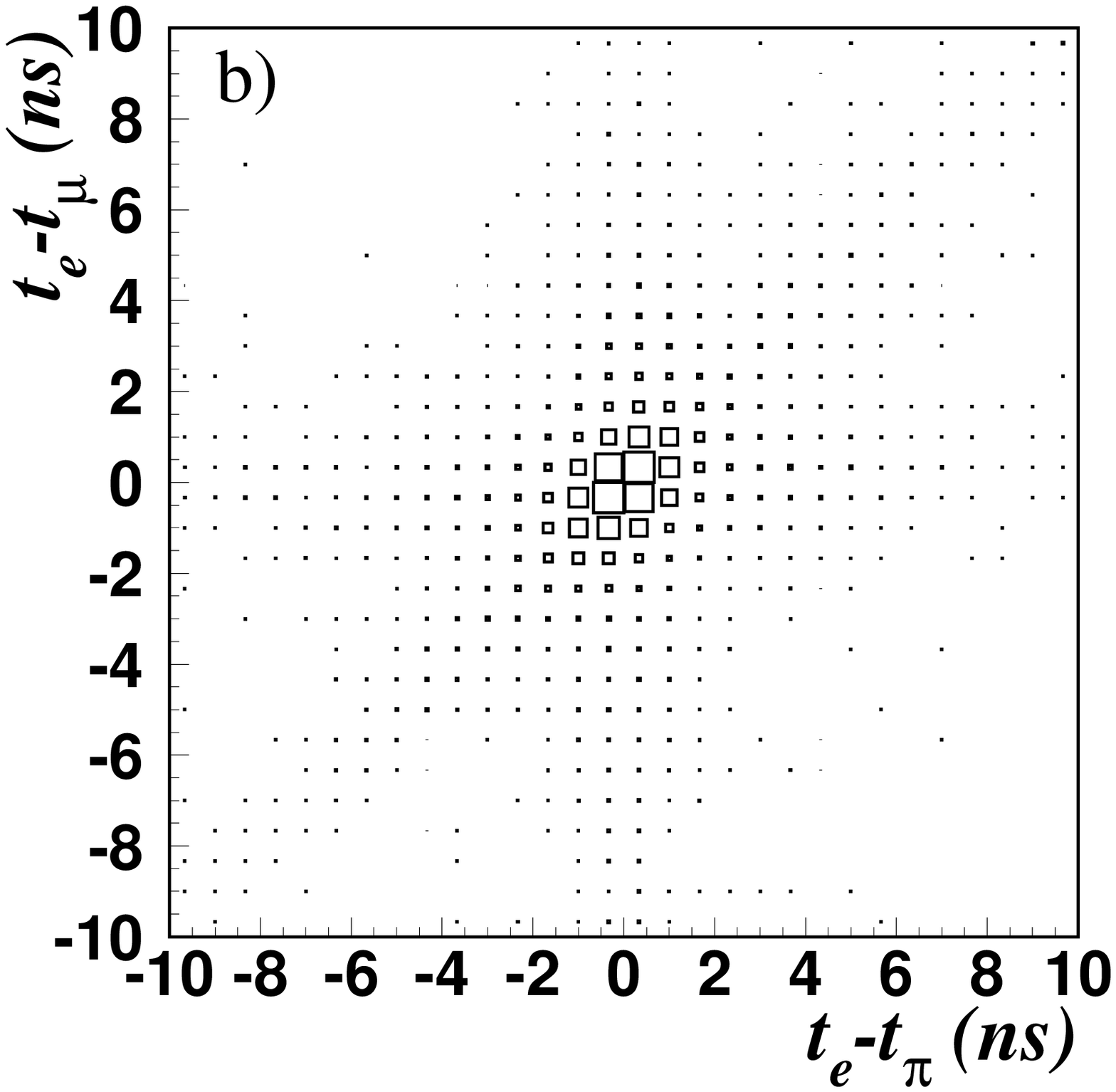}}
\resizebox{.25\textwidth}{!}{
\includegraphics{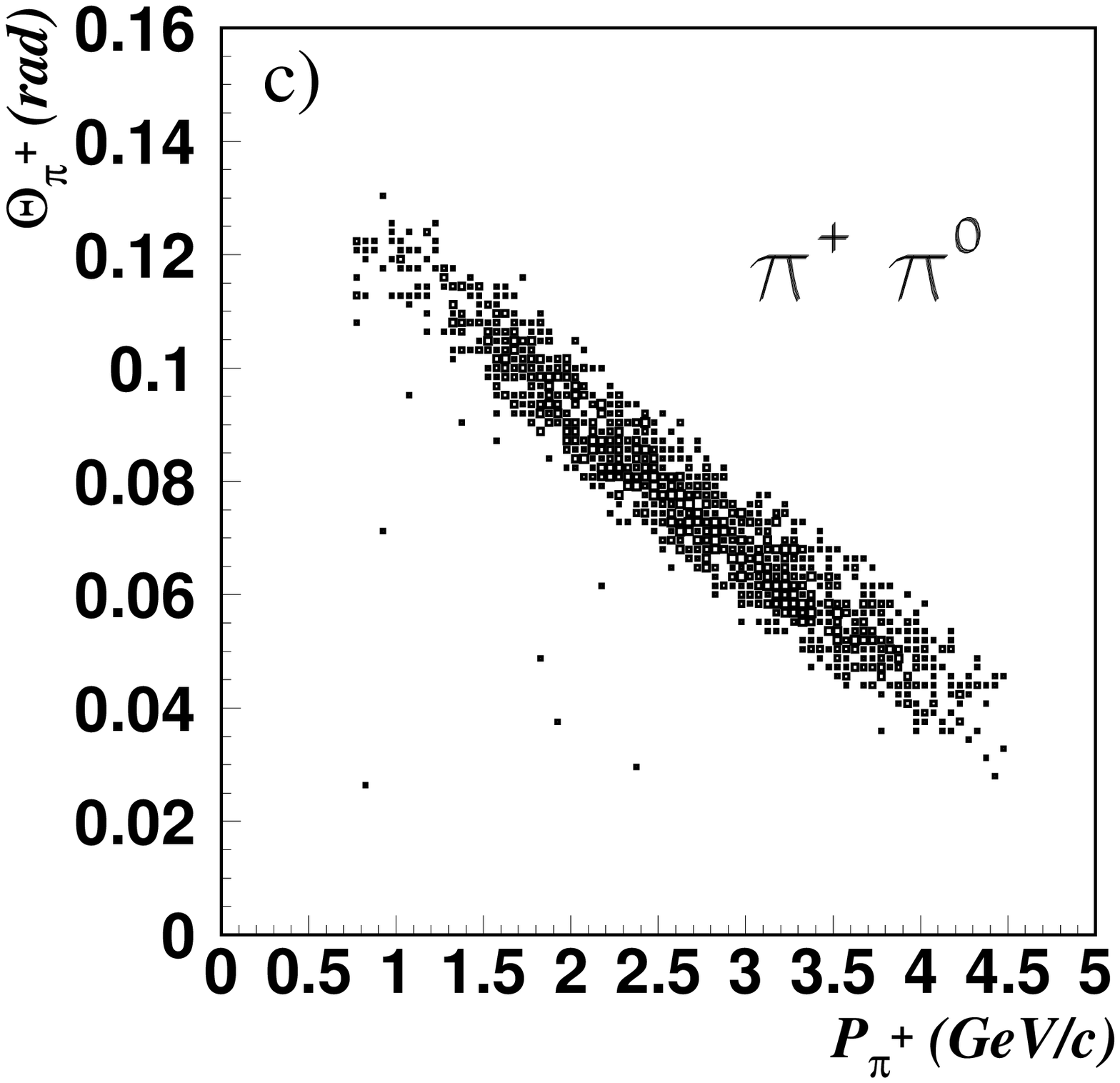}}
\resizebox{.25\textwidth}{!}{
\includegraphics{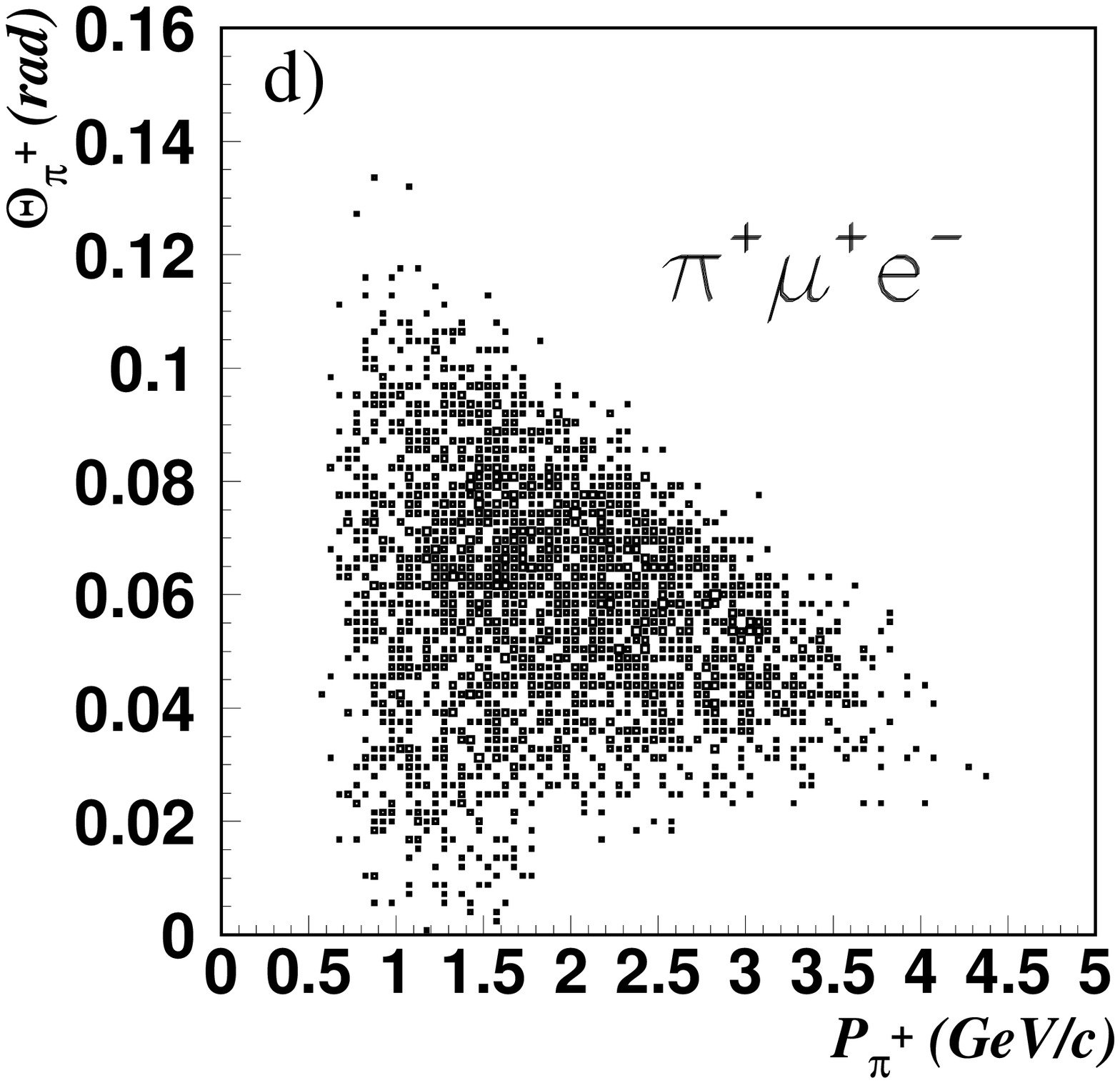}}}
\end{center}
\vspace*{-1cm}
\caption[ Scatter plot of differences between track times:
$t_{\pi^-}$ - $t_{\pi^+}$ versus $t_{\pi^-}$ - $t_{\pi^+}$]
{Scatter plot of differences between track times:
$t_{\pi^- (1)}$ - $t_{\pi^+ (2)}$ versus $t_{\pi^- (1)}$ - $t_{\pi^+ (3)}$
 for the high momentum accidental events [a)] and measured $K_{\tau}$
 events[b)];
Scatter plot of $\pi^+$ track angle at the vertex ($\theta$) versus
$\pi^+$ momentum ($P_{\pi}$) for the simulated $\kpi2$ [c)] and
 $\kpme$ [d)] events. 
}
\label{t_vs_t_tau_acc}
\end{figure*}
Kinematically,  the $K_{\tau}$ background
was suppressed by applying the invariant $M_{\pi\mu e}$ mass cut, since
assigning lower masses ($m_{e},m_{\mu}$) to the pions from the $K_{\tau}$ decay
effectively removed mass from the system as seen in  Fig. \ref{taubg_est}c. 
To estimate the level of the
$K_{\tau}$ background we used the TAU triggered data.
We selected $\pi^+\mu^+\pi^-$ events that could mimic $K_{\pi\mu e}$ (if
a $\pi^-$ was misidentified as an $e^-$)
 by applying the $K_{\pi\mu e}$
kinematic selection cuts and requiring identification of the $\pi^+$ and
$\mu^+$.
Events that were selected this way were
scaled by the TAU trigger hardware prescale factor ($10^4$) and a 
corresponding misidentification probability - P($\pi^-$ as $e^-$). 
To reduce the misidentification probability we constructed the log-likelihood
function for the purpose of $e^-$ identification. 
The likelihood included the number of p.e.
 and time registered by  C1L and C2L,
 the ratio E/P
 in the calorimeter and again the variable variable $E_{ratio}$.
The selected cut value for the $e^-$ PID likelihood, which resulted
 in the estimated misidentification
 probability P($\pi^-$ as $e^-$)=$(2.4\pm 1.4) \times 10^{-6}$,
ensured a satisfactory suppression
of the estimated $K_{\tau}$ background to the level of 0.1$\pm$0.1 events.

\subsection{Accidental background}

 To reject
accidental events effectively
 it was crucial to study and understand the precise
 accidental event mechanisms.
The simplest mechanism 
 would be a combination of two tracks 
from a real kaon decay and a single track from another decay, or beam
 background \emph{(2+1)}.
Another possibility is that all three tracks would originate from 
different decays or beam background \emph{(1+1+1)}.
To determine 
what particular combinations of accidental
tracks were dominant, the
time structure of accidental events was examined.
The time of each track ($t_{e},t_{\pi},t_{\mu}$) was defined as the 
average of the times registered by the
A counter and the calorimeter.
The $t_{e}-t_{\pi}$ and $t_{e}-t_{\mu}$
 track time differences were chosen to describe the
event timing.
A control sample of accidental three-track events was selected by 
requiring the reconstructed
kaon momentum to be greater
 then that of the beam ($P>$ 7 GeV/c) and removing
the $L_{Target}$ cut. To increase the sample size,
 cuts on the vertex quality
$S_{norm}$, and the invariant mass $M_{\pi\mu e}$, were
removed.
 Accidental events, selected in such a manner, showed a distinct {\em cross}
 in the scatter plot of the track time difference
as seen in Fig. \ref{t_vs_t_tau_acc}a,
indicating
that accidental events had two main components.
 The horizontal part of the cross identified
 the $\pi^+$ 
 as the  additional accidental track
 paired with the two tracks ($e^-\mu^+$) from another decay and the
 vertical part identified $\mu^+$ as an
 accidental track. Accidental
 $e^-$ tracks were not observed, since the scatter plot did not show a
linear correlation between the variables. Finally, 
triple coincidences \emph {(1+1+1)} 
appeared uniformly scattered and as we expected their contribution was
negligible.
 For the signal events, the correlation between the
 track time difference was 
examined using the measured response of the reconstructed $K_{\tau}$ events 
as can be seen in Fig. \ref{t_vs_t_tau_acc}b.

The difference in the time response between the signal and accidental events
 was
used to create a time quality estimator variable,
$T_{max}$, which 
was defined as the maximum absolute difference between the
track times $t_{\pi}$ or $t_{\mu}$ and the average times of the
two remaining tracks $t_{e\mu}$ or $t_{e\pi}$ respectively. 
 The ratio between the two accidental components
was estimated to be approximately one to two
(34\% $e\mu$+$\pi$ and 66\% $e\pi$+$\mu$).
The most reasonable candidate for the
two-track part ($e\pi$) of the main accidental component ($e\pi+\mu$)
was the $K_{\pi2}$ decay with a subsequent $\pi^0 \ra e^+e^-\gamma$ decay.
In the two-body decay $K_{\pi2}$ the pion momentum and angle of
the pion track at the vertex are correlated, as shown in Fig.
 \ref{t_vs_t_tau_acc}c. On the other hand three-body decay signal
 events do not show such a correlation  (Fig. \ref{t_vs_t_tau_acc}d).
The effect was best observed when the $\pi^+$ angle-momentum dependence was 
projected on the axis orthogonal to the axis
 of almost linear correlation between the angle and momentum of 
a $\pi^+$ from the $K_{\pi2}$ decay as shown in Fig. \ref{t_vs_t_tau_acc}c,
 equivalent to using the variable $z=\theta_{\pi}$/0.14+$P_{\pi}$/5
($P_{\pi}$ in GeV/c and $\theta$ in radians). 
By requiring the variable $z$ to be less then one, 
60\% of accidental background events were rejected 
with 80\% of them being from the $e\pi+\mu$ component.
This  caused only a 13.5\% loss in
 signal sensitivity.
\begin{figure*}
\centerline{
\resizebox{.25\textwidth}{!}{
\includegraphics{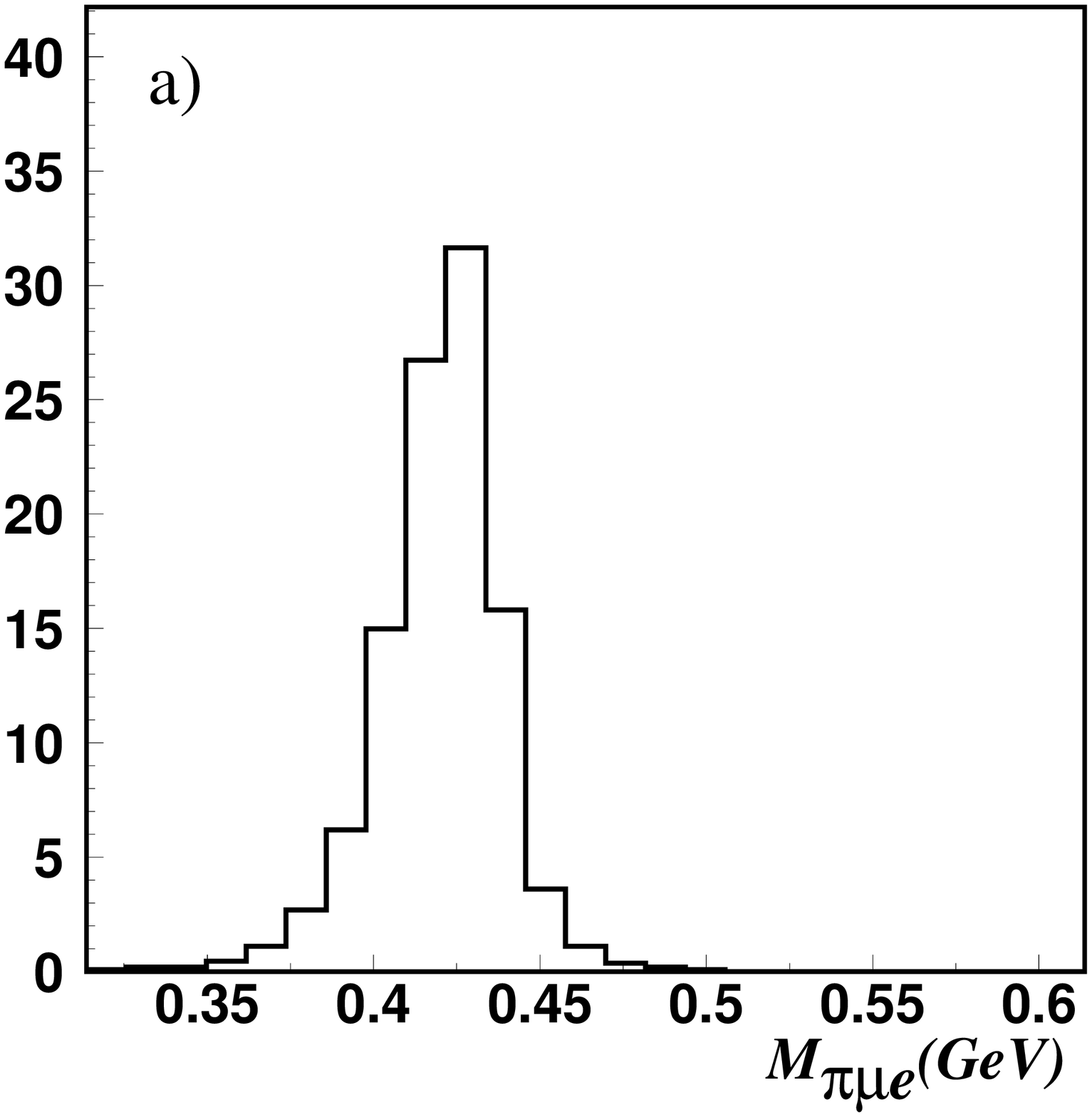}}
\hfill
\resizebox{.25\textwidth}{!}{
\includegraphics{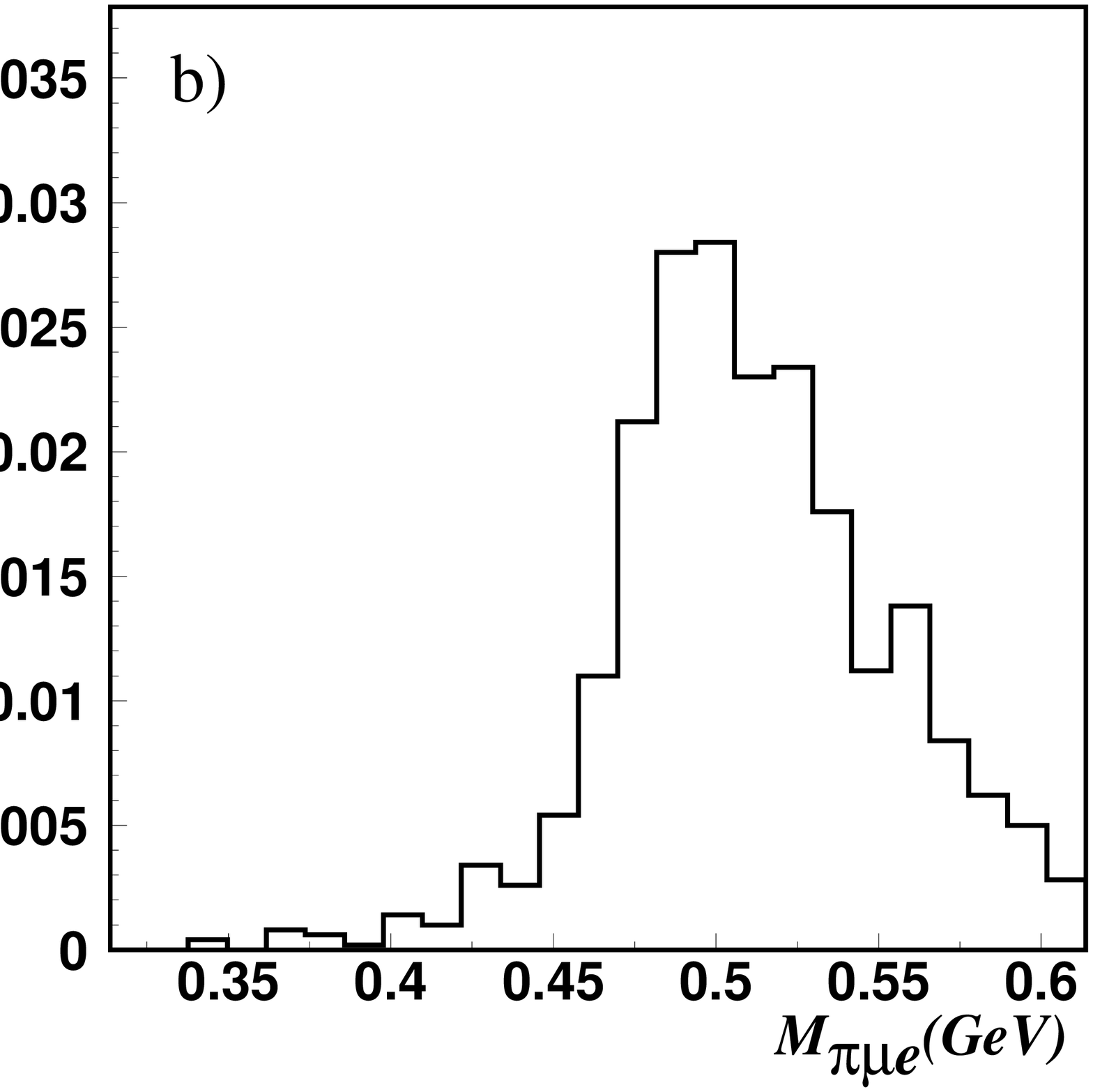}}
\hfill
\resizebox{.25\textwidth}{!}{
\includegraphics{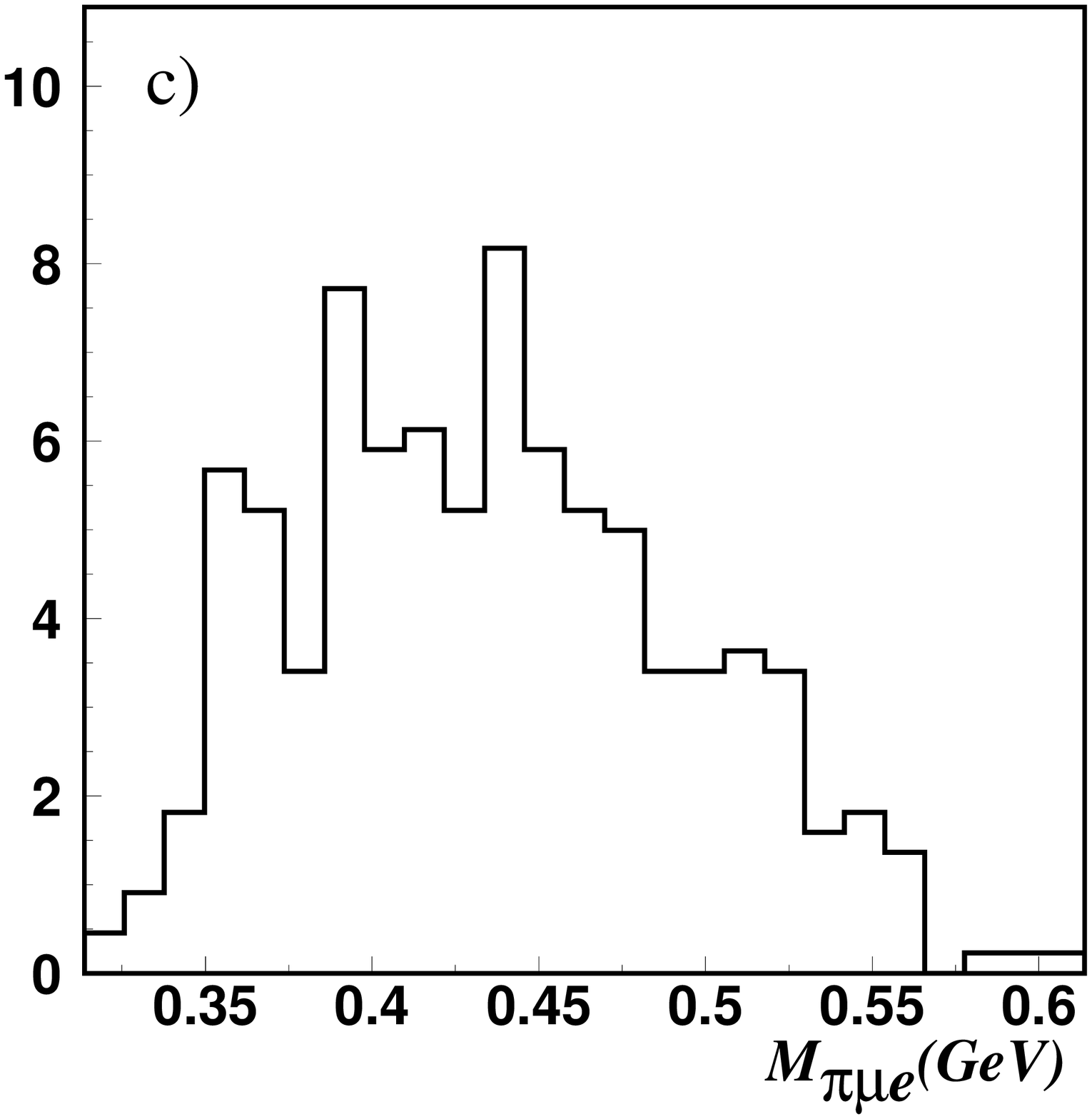}}
\hfill
\resizebox{.25\textwidth}{!}{
\includegraphics{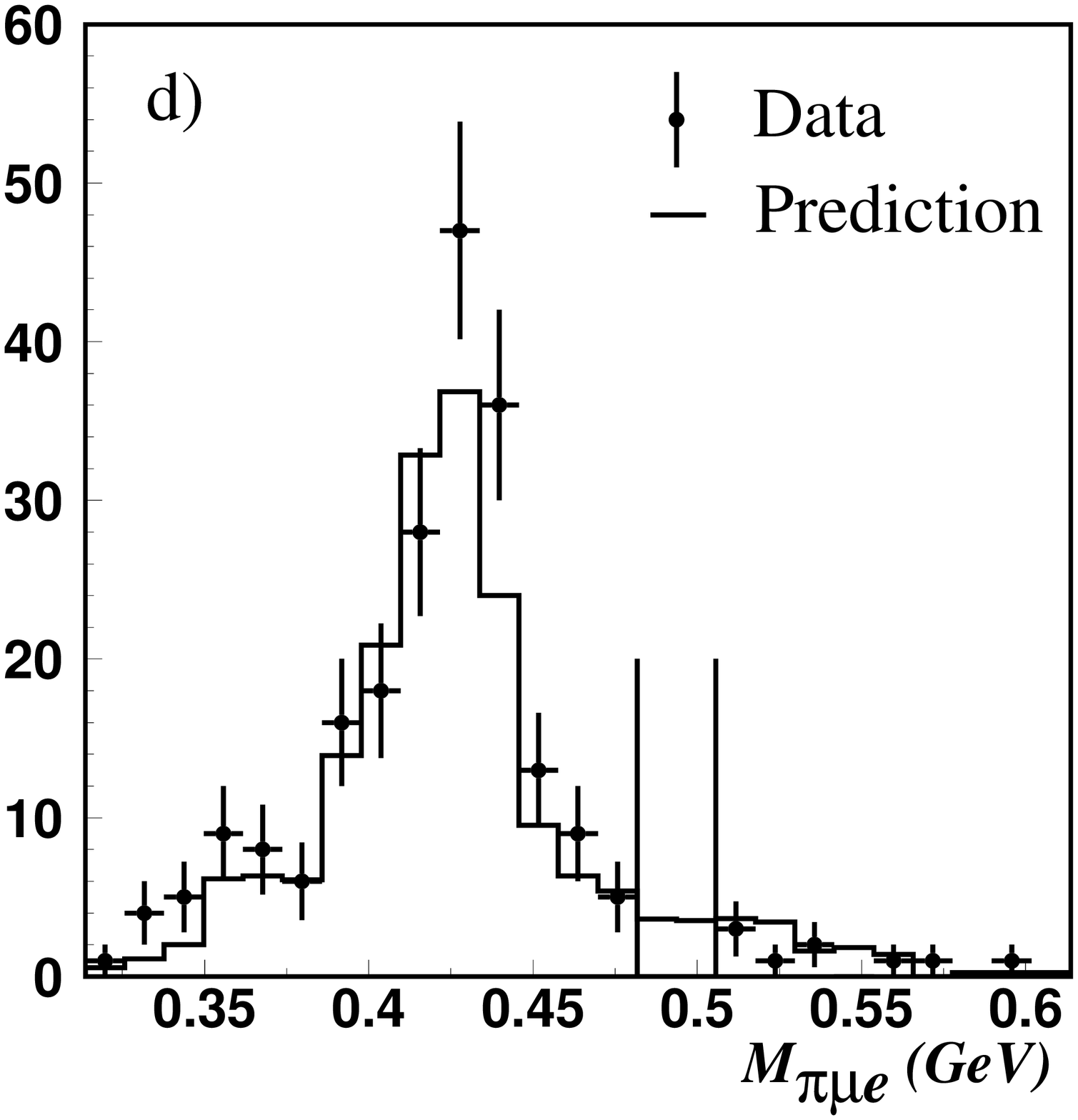}}}
\caption[The $M_{\pi\mu e}$ mass spectrum of the
predicted backgrounds]
{The reconstructed $M_{\pi \mu e}$ mass distribution for the:
a) estimated $K_{\tau}$ background; b) estimated $K_{Dal}$ background;
c) estimated accidental background; d) Comparison of the $M_{\pi\mu e}$ mass
distribution for the $K_{\pi\mu e}$ data (markers with error bars) and total
estimated background (histogram).}
\label{mpme_real_and_pred}
\end{figure*}

 The number of accidental background events was estimated using 
the number of observed out-of-time
$K_{\pi\mu e}$ candidates (3 ns $< T_{max} <$ 8 ns) which otherwise satisfied 
all $K_{\pi\mu e}$ cuts. 
To extrapolate into the acceptable in-time signal region, 
the timing distribution of the
high momentum ($P>6.9$ GeV/c) $K_{\pi\mu e}$ candidates was used.
From that timing distribution we determined the
projection factor $R$ (the ratio between the number of
high momentum $K_{\pi\mu e}$ candidates with $T_{max}<3$ ns and the ones with
the the $T_{max}$ in the range of 3 to 8 ns).
The expected level of accidental background in the signal region was thus
 determined to be $8.2 \pm 1.9$ events.
 Data study confirmed that
the projection factor $R$, used for accidental background estimate, 
did not depend on
 the momentum cut-off value, which was varied 
in the range of 6.9 to 7.7 GeV/c. To test the validity of
using high momentum accidentals to extract the time distribution of 
$K_{\pi\mu e}$ accidental background, we selected
a sample of $K_{\pi\mu e}$ candidates with the
 $M_{\pi\mu e}$ mass cut reversed and no timing ($T_{max}$) cut 
applied. 
 The timing response from the latter sample was a superposition 
of the timing response from $K_{\pi\mu e}$ accidental background 
and the timing response of the $K_{\tau}$ and
$K_{Dal}$ backgrounds. The tail ($T_{max}>3$ ns) of the timing spectrum,
 which was
dominated by accidental events in that sample,
 matched the timing distribution of the high momentum events.

\section{Likelihood analysis}
\subsection{Check of the background estimates}
Following the principles of the so called \emph{blind analysis} the
$K_{\pi\mu e}$
candidate events were examined, while the signal region was excluded
by reversing the cut on the invariant kaon mass
($|M_{\pi\mu e}-0.4937|>0.012\ \rm{GeV} $).
The invariant $M_{\pi\mu e}$ mass 
distribution for the 216 observed $K_{\pi\mu e}$ candidates
as illustrated in Fig. \ref{mpme_real_and_pred}d exhibited a peak 
at lower mass values, which was caused by the $K_{\tau}$ background. 
 In order to
check the validity of the background 
estimates,
the $M_{\pi\mu e}$ mass distribution for the $K_{\pi\mu e}$ candidates
 was compared
with that of the predicted backgrounds.
Following the same algorithm as that of the background estimates
 the $M_{\pi\mu e}$ distributions  were generated 
for the main background processes using measured data:
for the $K_{\tau}$ background, observed $\pi^-\pi^+\mu^+$ events
were scaled 
by the misidentification probability $P(\pi^-$ as $e^-)$.
For the $K_{Dal}$ background, observed $e^-e^+\mu^+$ events
were scaled
by the misidentification probability $P(e^+$ as $\pi^+)$.
Finally, for the accidental background, the
 out-of-time $K_{\pi\mu e}$ candidate events
were scaled
by the projection factor $R$ (ratio of in-time to out-of-time accidental
events).
 The satisfactory agreement between the
predicted and observed events as displayed in Fig. \ref{mpme_real_and_pred}d
 served as a consistency check for
our estimates of the background processes and
misidentification probabilities ($P(e^+$ as $\pi^+)$ and $P(\pi^-$ as $e^-)$).

\subsection{General $K_{\pi\mu e}$ likelihood}
The background study indicated that 98\% of the background to $K_{\pi\mu e}$ 
was due to 
accidental events.
As discussed above, rather than imposing
additional or tighter cuts, the likelihood function was chosen to
suppress this background.
The general $K_{\pi\mu e}$ likelihood
 was
constructed using 
the kinematic and timing variables which showed the most difference
in their response for signal and accidental events.
The following six variables were selected: target likelihood, $L_{Target}$, 
vertex quality, $S_{norm}$,
 reconstructed
$M_{\pi\mu e}$ mass, timing cut, $T_{max}$, sum
of the $\chi^2$ for the track reconstruction fits, and $T_{extra}$, the minimum
time difference between the
time of the extra clump in the calorimeter and time of the track pair, which
is closest in time. Data studies
 showed that within the statistical uncertainties the variables chosen 
were uncorrelated both for the signal and for the background.
This allowed us to calculate the general likelihood for $K_{\pi\mu e}$ and
 accidentals in the following manner:
\begin{equation}
{\cal{L}}_{\pi\mu e}(x_1...x_6)=\displaystyle \sum_{i=1}^{6}
{\log P_i^{\pi\mu e}(x_i)}
\end{equation}
\begin{equation}
{\cal{L}}_{Acc}(x_1...x_6)=\displaystyle \sum_{i=1}^{6}
{\log P_i^{Acc}(x_i)},
\end{equation}
\noindent
where $x_i=(L_{Target}$, $S_{norm}$, $M_{\pi\mu e}$, 
$\chi^2_{Track}$,
 $T_{max}$, $T_{extra}$) and $P_i^{\pi\mu e}$, $P_i^{Acc}$ are 
 one dimensional 
PDFs for signal and background, respectively. For the signal,
only the $M_{\pi\mu e}$ mass PDF was generated from the Monte Carlo simulated
events with remaining PDFs generated from the measured data by using
reconstructed $K_{\tau}$ events. For the accidental events, respective 
PDFs were generated
from the data, using out-of-time 
$K_{\pi\mu e}$ candidates, except for $T_{max}$ and $T_{extra}$ for which
the PDFs
were generated using high momentum $K_{\pi\mu e}$ candidates\cite{sher04}.
 The total relative likelihood ${\cal{L}}$ for $K_{\pi\mu e}$ was defined 
 as the difference between ${\cal{L}}_{\pi\mu e}(x_1...x_6)$ and
${\cal{L}}_{Acc}(x_1...x_6)$.

The likelihood distribution for signal events 
was generated using the measured data. With the exception of the
invariant $M_{\pi\mu e}$ mass, the response of the $K_{\tau}$ events for
the likelihood variables was identical to that of the signal.
 The $M_{\pi\pi\pi}$ mass resolution of $K_{\tau}$ events had to be scaled
by a factor of 1.725 in order to match the mass resolution of 
$K_{\pi\mu e}$ events. The latter was
 determined from the Monte-Carlo simulations\cite{sher04}. 
Consequently, the signal likelihood template was generated using
the good $K_{\tau}$ events with the scaled $M_{\pi\pi\pi}$ mass
used as the $M_{\pi\mu e}$ mass.

The likelihood distribution for accidental events was obtained using
a sample of out-of-time $K_{\pi\mu e}$ candidate events.
Since all six variables used in the likelihood had to be
within the final cut values, 
the timing response was generated from 
a library of the measured timing response of
 the high momentum ($P>7$ GeV/c) accidental 
 events\cite{sher04}.
\begin{figure}
\begin{center}
\centerline{
\resizebox{.34\textwidth}{!}{
\includegraphics{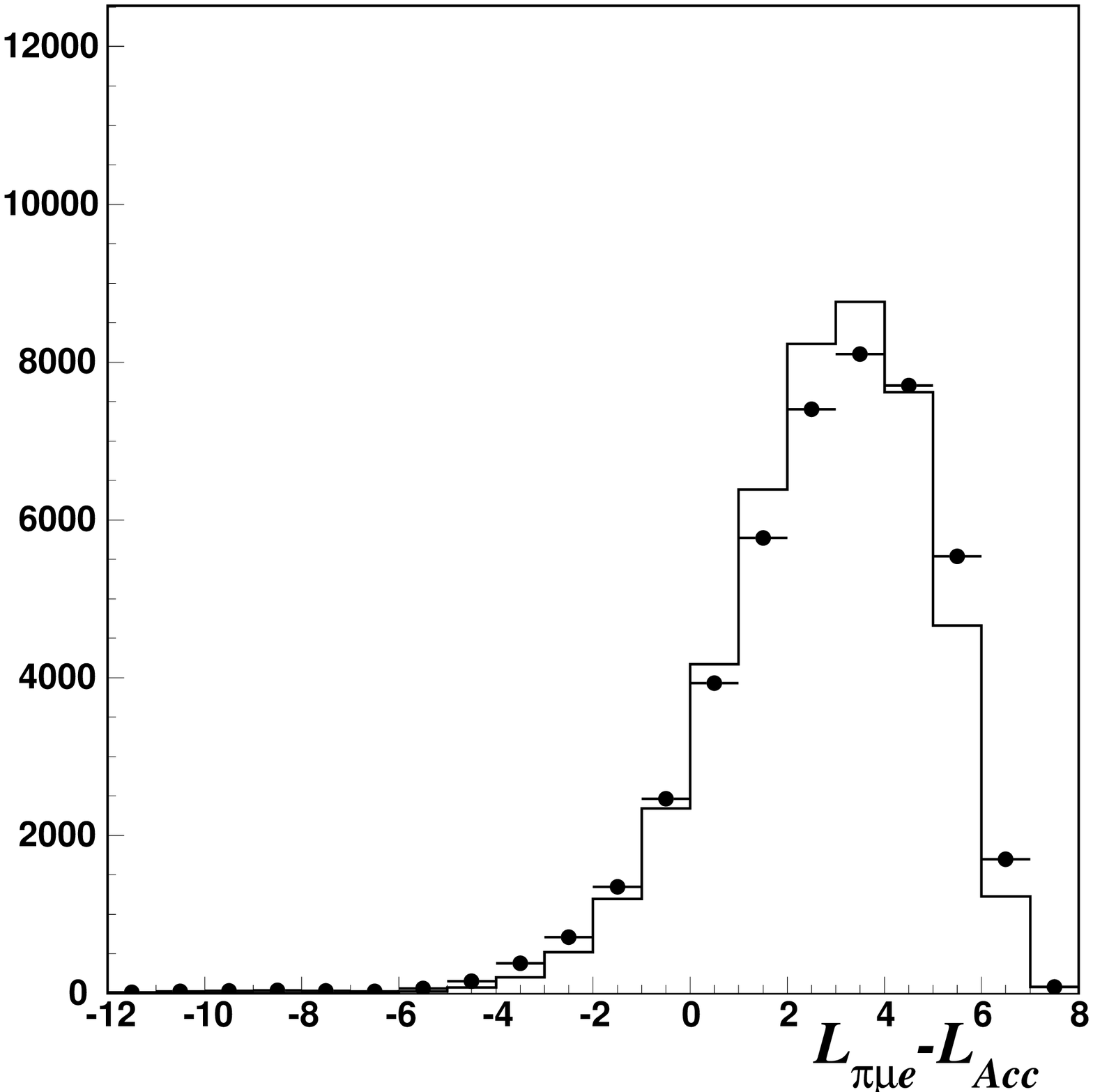}}}
\centerline{
\resizebox{.34\textwidth}{!}{
\includegraphics{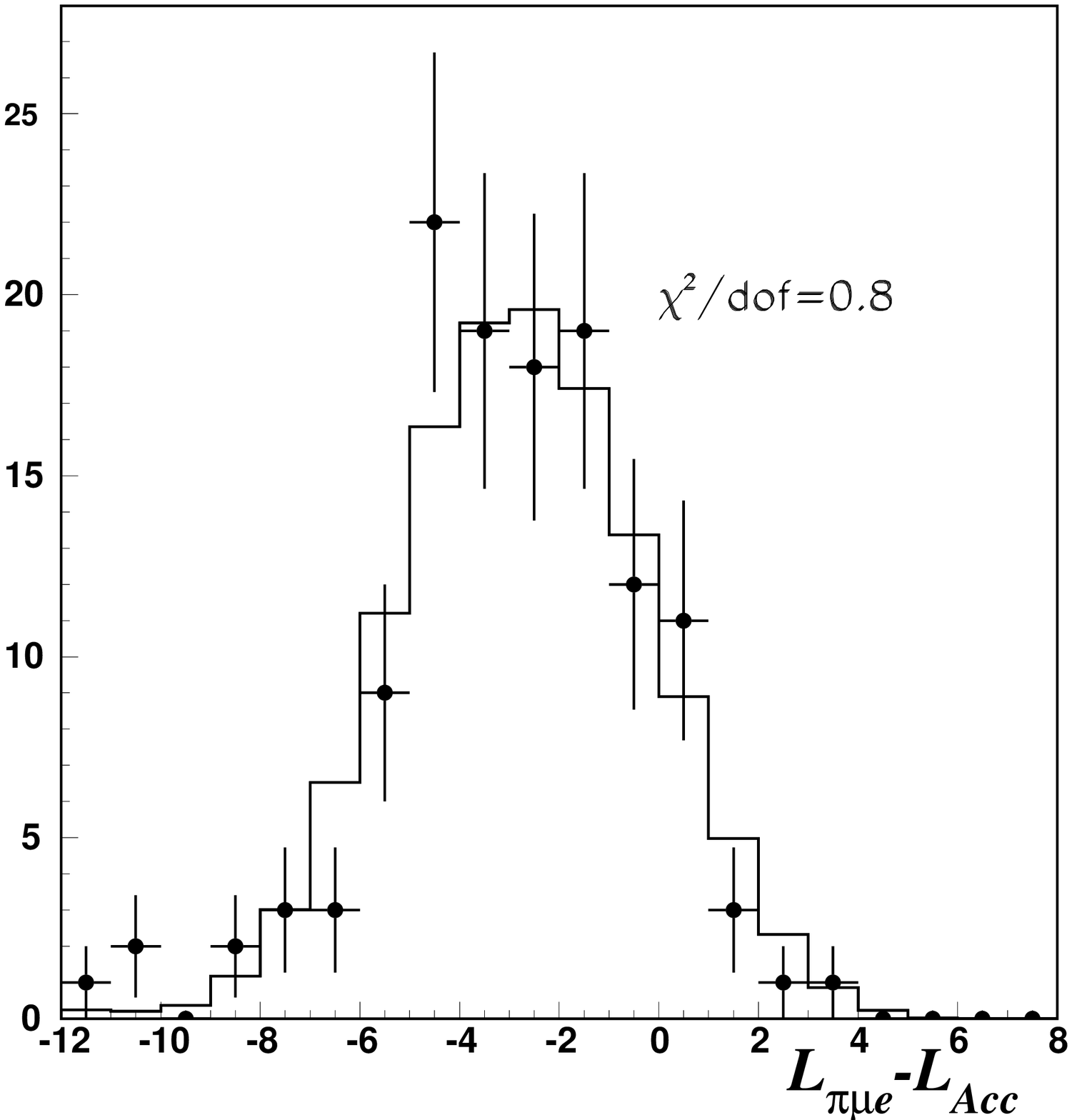}}}
\end{center}
\vspace*{-1cm}
\caption[General likelihood for the simulated signal and background]
{ 
Comparison of the general likelihood
distributions for measured (markers with error bars) and simulated 
directly from the PDFs events 
 for the $K_{\pi\mu e}$ (top) and the accidental background (bottom).
The $K_{\pi\mu e}$ likelihood was calculated from the
 measured $K_{\tau}$ event sample (see text for details).}
\label{glik_sim_comp}
\end{figure}
To test the assumption about the independence of the variables used in the
general likelihood,
their response was simulated directly from the corresponding PDFs.
The fact that the simulated likelihood distributions compared
favorably 
with the ones derived
 from the data (see Fig. \ref{glik_sim_comp}), confirmed our
study of variables correlation and ensured the validity of the likelihood
factorization of Eq. (\ref{eq:lik_fact}).

\subsection{Determining the number of $K_{\pi\mu e}$ events}
\label{sec:bayes}
\subsubsection{The Bayesian approach}

To determine the number of signal events, in the presence of  
a significant number of accidental background events (8.2 $\pm$ 1.9),
 the Bayesian approach\cite{Pdg04} was used. 
The likelihood function for Poisson-distributed 
data was defined as follows:
\begin{equation}
L_{P}(\vec{m};\vec{n}) = \displaystyle \prod_{i=1}^{k}
\frac { m_{i}^{n_{i}} e^{-m_{i}} }{  n_{i}! }
\label{eq:poisson_lik}
\end{equation}
where
\begin{eqnarray}
k       & = & \mathrm{arbitrarily\ chosen\ number\ of\ bins} \nonumber \\
n_{i}   & = & \mathrm{the\ number\ of\ observed\ events\ in\ the}\ i_{th}\
              \mathrm{bin} \nonumber \\
\vec{n} & = & (n_{1},n_{2},...,n_{k}) \nonumber \\
N       & = & \mathrm{total\ number\ of\ observed\ events}\
( = \displaystyle \sum_{i=1}^{k} n_{i}) \nonumber \\
m_{i}   & = & \mathrm{the\ number\ of\ expected\ events\ in\ the}\ i_{th}\
              \mathrm{bin} \nonumber \\
\vec{m} & = & (m_{1},m_{2},...,m_{k}) \nonumber
\end{eqnarray}

The number of expected events in the $i_{th}$ bin 
of the likelihood distribution was parameterized in the following way:
\begin{equation}
m_{i} = N_{sig} P_{\pi \mu e_{i}} + N_{bg} P_{acc_{i}} 
\label{eq:mi_param}
\end{equation}
where $N_{sig}$ was the number of $K_{\pi\mu e}$ events, $N_{bg}$ was
the number of accidental events, and $P_{\pi\mu e_{i}}$ and $P_{acc_{i}}$ were
the probabilities to be in the $i_{th}$ bin for the signal and 
the accidental background,
 respectively.
The parameter $N_{bg}$ was determined earlier
 from the study of accidental background.
 The probabilities ($P_{\pi\mu e_{i}}$ and $P_{acc_{i}}$)
 were extracted using the 
 likelihood templates, created
for the $K_{\pi\mu e}$ and accidental hypothesis.
The statistical uncertainty of the background estimate was accounted 
for by adding a Gaussian factor to Eq.(\ref{eq:poisson_lik}):
\begin{equation}
L_{P} = \displaystyle \prod_{i=1}^{k}
\frac { m_{i}^{n_{i}} e^{-m_{i}} }{  n_{i}! } \times
 {\rm{exp}}{ \left( -\frac{(N_{bg}-N^{(0)}_{bg})^2}{2\sigma_{bg}^2} \right)}
\label{eq:poisson_lik1}
\end{equation}
where\\
\begin{center}
$N_{bg}^{(0)} = 8.2$ and  $\sigma_{bg}=1.9$\\
\end{center} 
 Under the parameterization
of $m_i$, given in Eq. (\ref{eq:mi_param}),
 the Poisson-distributed data likelihood function, Eq. (\ref{eq:poisson_lik1}),
was transformed: $L_{P}(\vec{m};\vec{n})$ $ \Rightarrow$ $L(N_{sig},N_{bg})$.
 As the next step, the probability was normalized to unity:
\begin{equation}
P(N_{sig},N_{bg})=\frac{ L(N_{sig},N_{bg})}
{\int_{0}^{\infty} dN_{bg} \int_{0}^{\infty} L(N_{sig},N_{bg})dN_{sig}}
\end{equation}
To determine the upper limit on the number of signal events at the
90 $\%$ C.L. one integrates the 
probability $P(N_{sig},N_{bg})$ from
zero up to 
$N_{sig}^{max}$, until the 0.9 probability level is reached:
\begin{equation}
\frac{\int_{0}^{\infty} dN_{bg}
 \int_{0}^{N_{sig}^{max}} L(N_{sig},N_{bg})dN_{sig}}
{\int_{0}^{\infty} dN_{bg} \int_{0}^{\infty} L(N_{sig},N_{bg})dN_{sig}} = 0.90
\label{eq:integral}
\end{equation}

By solving  Eq. (\ref{eq:integral}) 
one can numerically determine the number of signal events at the 90$\%$ C.L.
\subsubsection{Signal region}
\begin{figure}
\begin{center}
\centerline{
\resizebox{.42\textwidth}{!}{
\includegraphics{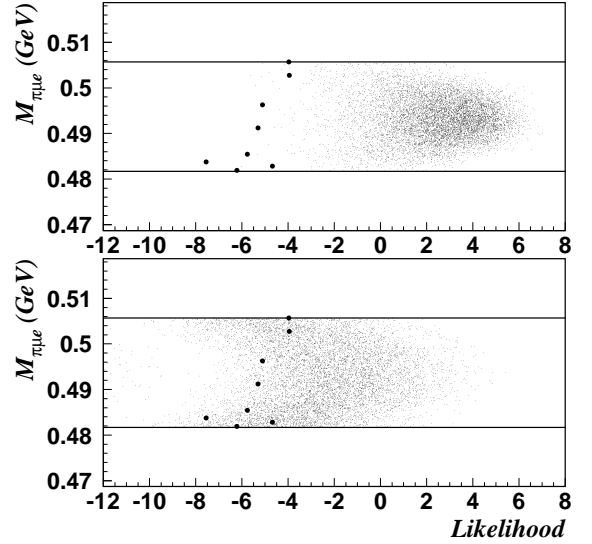}}}
\end{center}
\vspace*{-1cm}
\caption[General likelihood for the simulated signal and background]
{ 
Scatter plot of $K_{\pi\mu e}$ data (bold dots) and simulated
signal(top) and background (bottom). The abscissa is the log-likelihood of
the reconstructed events, the ordinate is the invariant mass ($M_{\pi\mu e}$)
of the detected particles. The horizontal lines mark the 3$\sigma$ mass
 region.}
\label{pme_acc_observed}
\end{figure}

Eight events survived the final $K_{\pi\mu e}$ selection,
 which was consistent with our background prediction of 8.2 $\pm$ 1.9. 
 Examination of
 the likelihood distribution of the surviving events demonstrates that they are
clearly not consistent with the signal hypothesis
(see Fig. \ref{pme_acc_observed}). 
By solving Eq. (\ref{eq:integral}) for the surviving events
an upper limit on
the number
of $K_{\pi\mu e}$
 events at a 90\% C.L. was determined to be $N_{sig}<$ 2.4.

A simple cut
 on the likelihood would remove all surviving events with only a 
3$\%$ loss in efficiency. 

\begin{figure}

\begin{center}
\centerline{
\resizebox{.5\textwidth}{!}{
\includegraphics{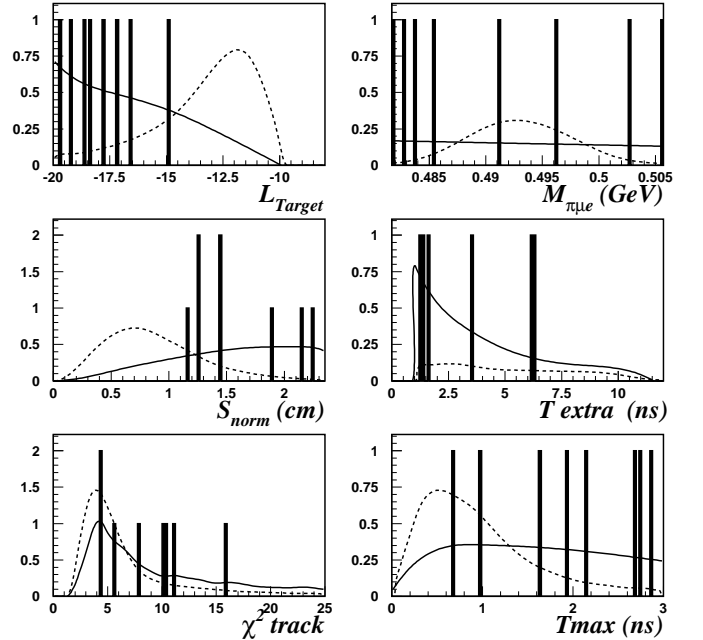}}}
\end{center}
\vspace*{-1cm}
\caption[Distributions of the likelihood variables for observed events]
{Distributions of target likelihood ($L_{Target}$),
 reconstructed kaon mass $M_{\pi\mu e}$, 
vertex quality $S_{norm}$, minimum time difference between extra clumps
in the calorimeter and closest in time track pair ($T_{extra}$),
 sum of track's $\chi^2$,
 and $T_{max}$ for the $K_{\pi\mu e}$ data. For the reference purpose
 corresponding PDFs 
are shown for signal (dotted line) and background (solid line) hypotheses.}
\label{final_dist}
\end{figure}

The distributions of the six variables used in
 the general likelihood 
for the observed eight events 
along with the respective signal and background PDFs 
are presented in Fig. \ref{final_dist}.

\section{Normalization and Acceptance Calculation}
To calculate 
the $K_{\pi\mu e}$ branching ratio,
 the total number of $K^+$ 
decays had to be determined.
The total number of $K^+$ decays can
be determined indirectly by normalization to
a decay mode that is well understood and observable
in the detector, and has
 a relatively high branching ratio.
We used the $K_{\tau}$ decay as the normalization mode. The upper limit
on the branching ratio was determined according to the following formula:

\begin{equation}
B(\pi \mu e)<B(\pi\pi\pi) \cdot
\frac{\emph{$N(\pi \mu e)$}}{\emph{$N(\pi\pi\pi)$}}\cdot
\frac{\emph{$Accep(\pi\pi\pi)$}}{\emph{$Accep(\pi \mu e)$}} \cdot
C
\label{eq:upper_limit_mc}
\end{equation}             
where $B$ denotes the branching ratio of the decay, $N(\pi\mu e)$ is the 90$\%$
 C.L. upper limit on the number of signal events, 
$N(\pi\pi\pi)$ is the number of observed $K_{\tau}$ events,
 adjusted for prescale factors, $Accep$ is acceptance of the detector system,
and finally C is the correction factor accounting for efficiency differences 
between the selection of signal ($K_{\pi\mu e}$)
 and normalizer ($K_{\tau}$) decay modes (see Sec. \ref{sec:corf}).

\subsection{Total number of $K_{\tau}$ events}
\label{sec:tau_total}
As mentioned in Sec. \ref{sec:trigger}, we collected $K_{\tau}$ decays
in a minimum bias TAU trigger concurrently with $K_{\pi\mu e}$ events.
This allowed us to deduce  the number of normalization $K_{\tau}$ events
by analyzing the TAU triggered data.
To reduce the systematic 
uncertainties in the acceptance ratio calculation, the normalization
sample was kinematically selected using cuts which 
 were identical to the ones for $K_{\pi\mu e}$.
 However, some of the cuts used for the $K_{\pi\mu e}$ selection, 
including the PID cuts, the invariant $M_{ee}$ mass cut, and a cut ($z<1$) that
rejected $\pi^+$'s originating from $K_{\pi2}$ decays,
 were not applied.

\begin{table}
\caption [Cuts used to select normalization sample of $\kppp$ events]
{
Cuts used for determining the number of $K_{\tau}$ decays.
}
\begin{ruledtabular}
\begin{tabular}{llc}
 {\it{Variable}} &  {\it {Description}}     & {\it{Efficiency}}         \\[0mm]
\hline
$S_{norm}$ & vertex quality & $0.927 \pm 0.005$ \\[2mm]
 $\displaystyle \sum_{i=1}^{3}$
$\chi^{2}_{trk_{i}}$ 
 & sum of track's fit $\chi^2$  & $0.967 \pm 0.005$  \\[2mm]
$L_{Target}$  & target likelihood & $0.923 \pm 0.005$  \\[2mm] 
$T_{max}$  & time quality indicator & $0.911 \pm 0.004 $  \\[2mm] 
$T_{extra}$  & no extra clumps in the & 0.956 $\pm$ 0.005 \\
  & calorimeter within 1 ns &   \\[2mm]
$N_{\gamma}$ & photons veto & $0.965 \pm 0.005$ \\[2mm] 
$S_{2trk}$ & two track vertex quality & $0.959 \pm 0.005$ \\[2mm]

$M_{\pi\pi\pi}$      & reconstructed kaon mass  
 & $0.976 \pm 0.005$ \\ 
\end{tabular}
\end{ruledtabular}

\label{tab:tau_pass3_cuts}
\end{table}

\begin{figure*}
\centerline{
\resizebox{.34\textwidth}{!}{
\includegraphics{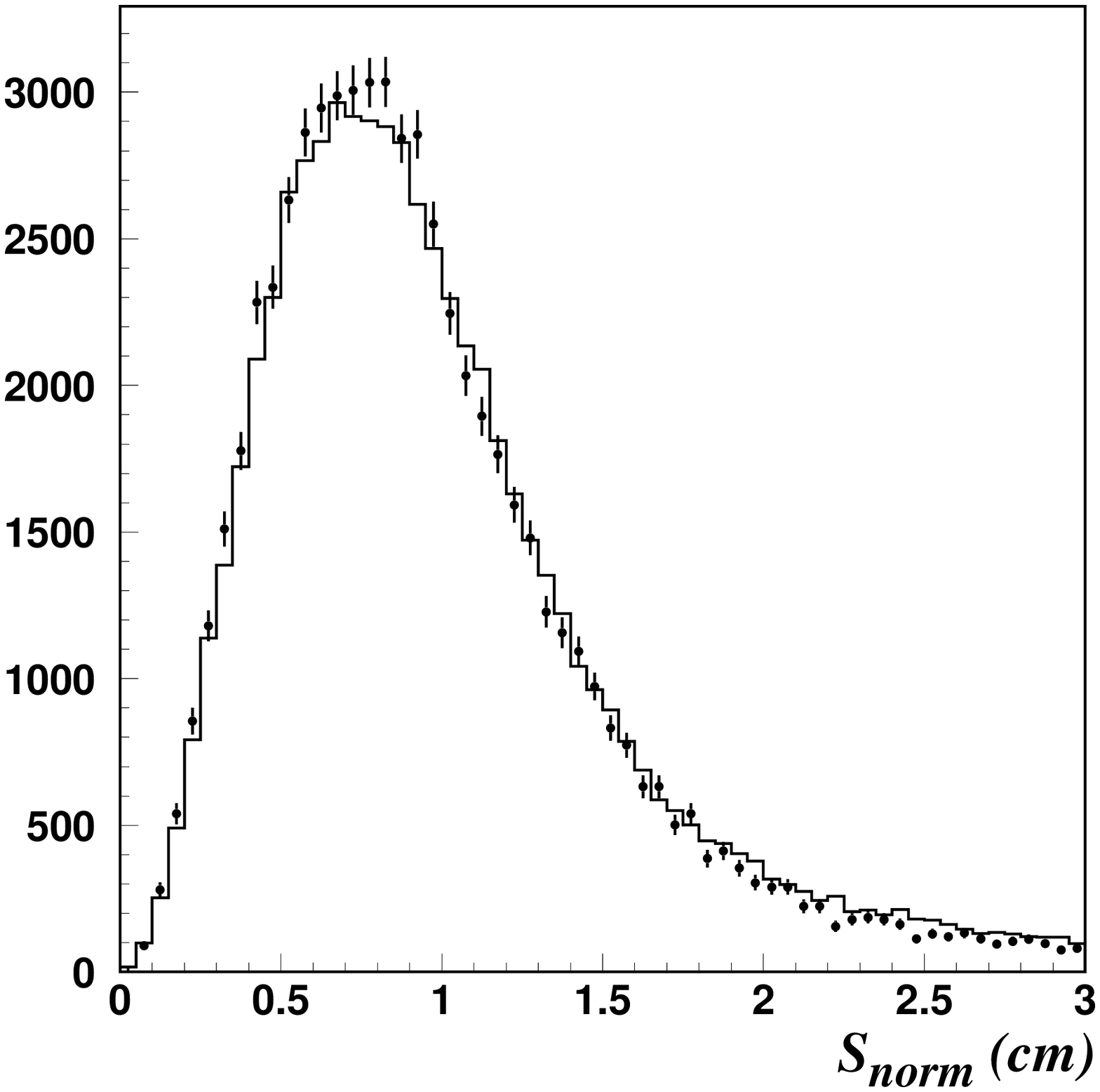}}
\hfill
\resizebox{.34\textwidth}{!}{
\includegraphics{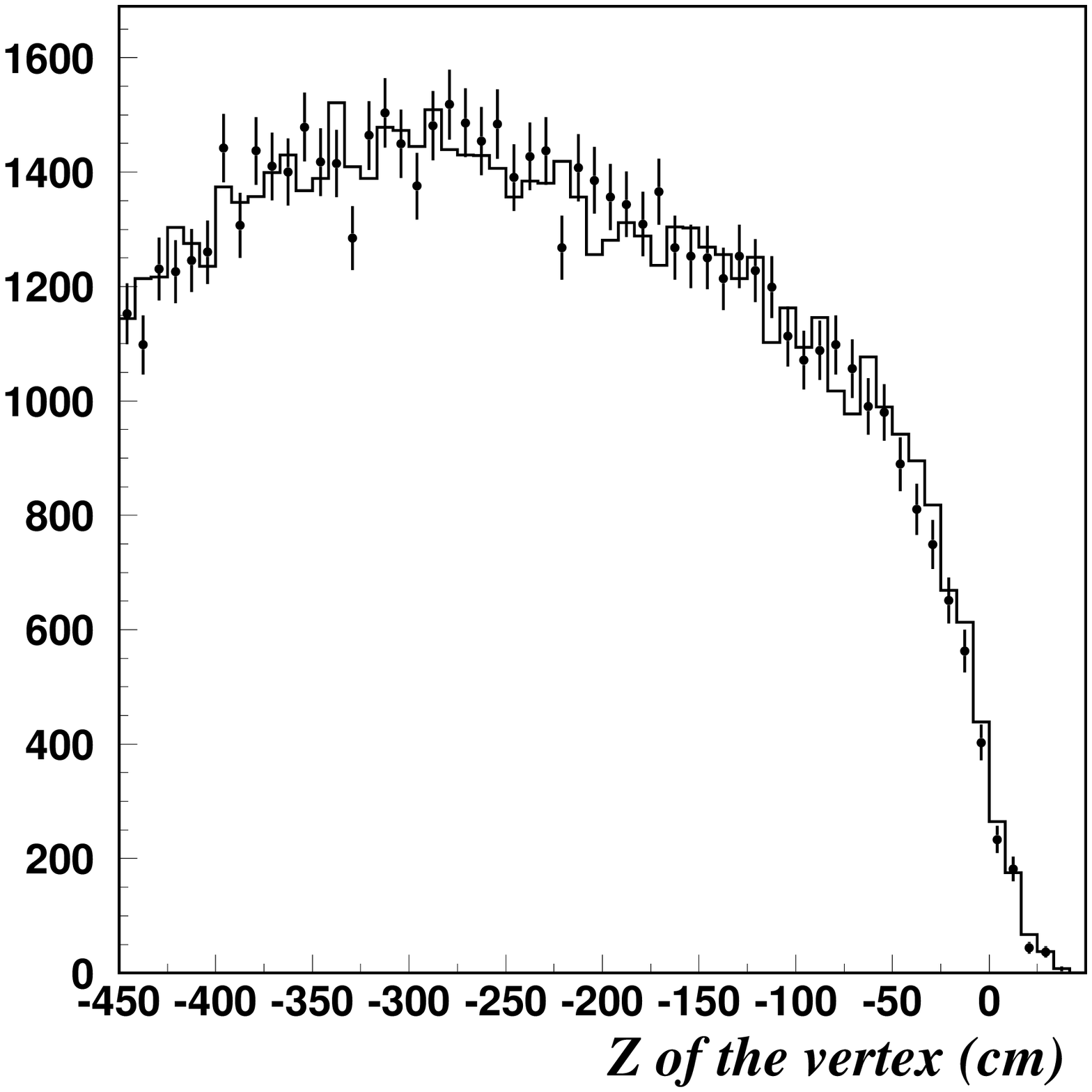}}
\hfill
\resizebox{.34\textwidth}{!}{
\includegraphics{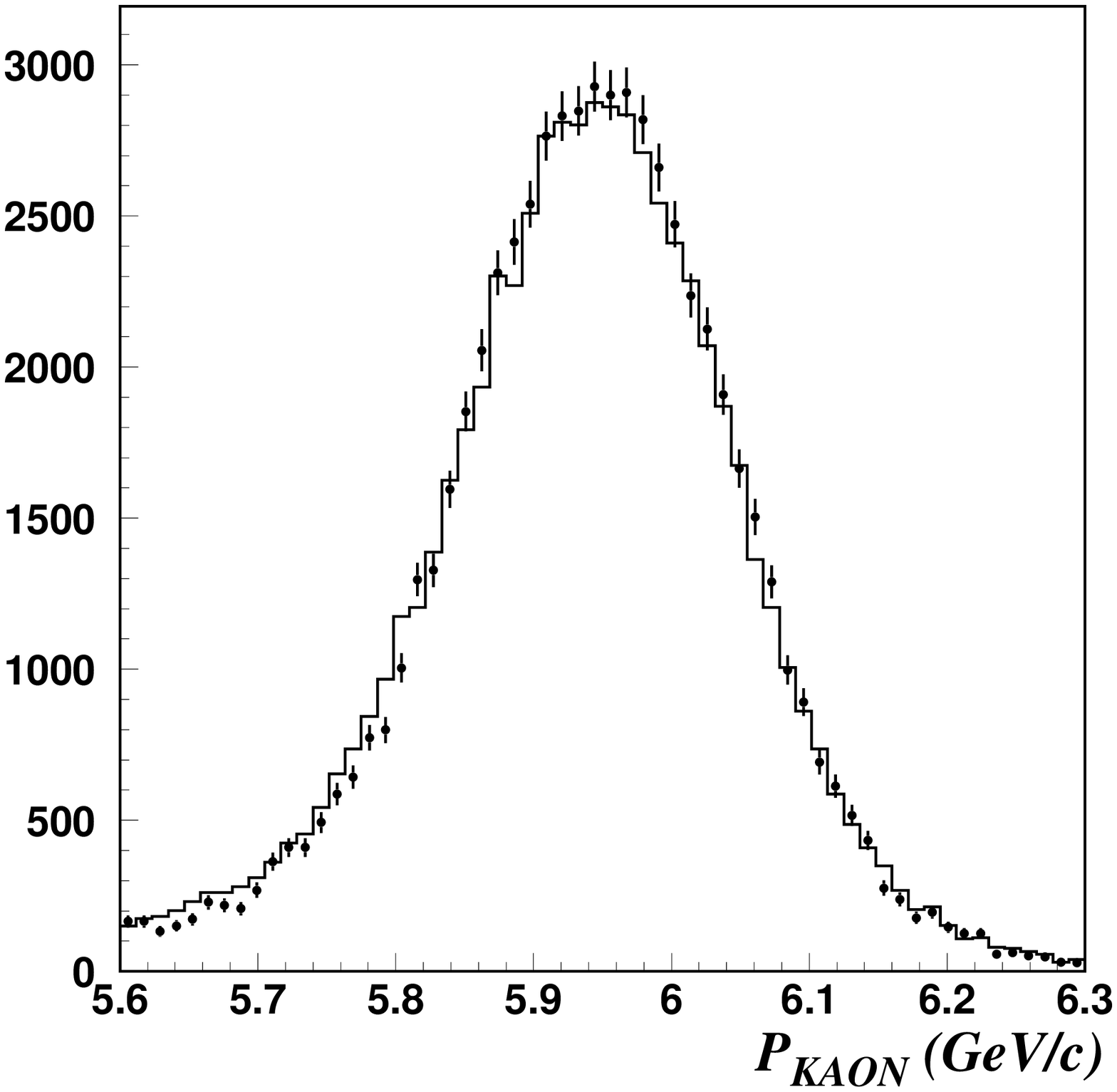}}}
\caption
{Comparison of the Monte Carlo simulations of $K_{\tau}$ decays 
(histogram) with data 
(markers with error bars). Left: distance of closest approach $S_{norm}$
to the common vertex for the three charged tracks; center: distribution
of decay vertices along the beam direction $Z$ ($Z=0$ at the entrance of
the first dipole magnet); right: Reconstructed $K^+$ momentum.}
\label{mcdata}
\end{figure*} 
The application of the selection cuts (see Table \ref{tab:tau_pass3_cuts})
 yielded 77226 events.
The
 total number of the $K_{\tau}$ events was determined after
accounting for the hardware ($10^4$) and software (50) prescale factors:
\begin{equation}
N(\pi\pi\pi)=77226 \times 50 \times 10^4 = 3.86 \times 10^{10}\ \rm{events}.
\end{equation}

\subsection {Monte Carlo simulations}

The simulation started with the kaon beam at the upstream end of
the decay region. The phase space of the beam was deduced from a
large sample of $K_{\tau}$ events, for which the incident $K^+$ could
be fully reconstructed. The $K^+$ was then allowed to decay in a
preselected mode along its trajectory in the decay region.
The interactions of the beam particles and the
 decay products with the detector
 were simulated using the 
{\tt {GEANT3}}\cite{GEANT}
 software package from CERN. The simulated detector response
was then analyzed using the same
reconstruction and selection procedures as data events.

In order to check the quality of the  simulation,
control
 distributions for various kinematic variables
were compared to measured data for $K_{\tau}$ and $K_{Dal}$ decays.
 Due to the different phase space and different
daughter particles,
 those two decays have significantly different control distributions.
The quality of the simulation is demonstrated in Fig. \ref{mcdata},
which displays the vertex quality, $S_{norm}$, which is a crucial parameter for
event reconstruction; the position of
the vertex along the direction of the beam, $Z$ of the vertex, which depends
on the detector acceptance; and reconstructed kaon
momentum, $P_{KAON}$, which is sensitive to the resolution.

As an additional consistency check, we estimated the number of
observed $K_{\tau}$
 decays, as registered by the TAU trigger, from the number
 of observed low $M_{ee}$ mass $K_{Dal}$ events, as
 registered by the EEPS
trigger, by applying
the acceptance ratio obtained from the Monte-Carlo simulations.
For the calculation of the low $M_{ee}$ mass
 $K_{Dal}$ events acceptance, we simulated
 $\kpi2$, $\kmu3$, $\kaye3$, $K^+ \rightarrow \pi^+ \pi^0 \pi^0 $ decays 
(all with a subsequent decay of a  $\pi^0 \rightarrow e^+e^-\gamma$)
 and weighted them by their measured branching ratios\cite{Pdg04}. The 
predicted number of $K_{\tau}$ decays 
 was consistent within the statistical uncertainty (1.5\%) 
with the number of observed $K_{\tau}$ events\cite{sher04}.
This underlined our understanding of the geometrical acceptance and
the efficiency of the various detector elements.

For $2.5 \times 10^5$ simulated $K_{\tau}$ events 31600 were accepted, 
resulting in:
\begin{equation}
{Accep(\pi\pi\pi)} = 0.1264 \pm 0.0008
\end{equation}

For $2.5\times10^5$ simulated $K_{\pi\mu e}$ events 17003  were accepted,
 resulting in:
\begin{equation}
{Accep(\pi\mu e)} = 0.0680 \pm 0.0005
\end{equation}
and
\begin{equation}
\frac{{Accep(\pi\pi\pi)}}{{Accep(\pi\mu e)}} = 1.86 \pm 0.02.
\end{equation}

         To estimate the influence of alignment inaccuracies
 we varied the size of a dead (desensitized) region where 
the beam passed
 (see Sec.\ref{sec:detector}), which
was the most sensitive parameter for both $K_{\pi\mu e}$ and $K_{\tau}$ decays
acceptance.
 We considered three
cases, for which we defined a smaller ($\Delta x=
\Delta y=$-0.5cm), nominal, and a larger dead region ($\Delta x=
\Delta y=$+0.5cm).
 While the changes in the acceptance
 of the $K_{\tau}$ and $K_{\pi\mu e}$ decays were noticeable,
the acceptance ratio showed the following variation:
\begin{itemize}
\item{ 1.88 $\pm$ 0.02 - smaller dead region;}
\item{ 1.86 $\pm$ 0.02 - nominal dead region;}
\item{ 1.83 $\pm$ 0.02 - larger dead region.}
\end{itemize}

From this study we deduced a systematic uncertainty of roughly 1.5\% on
the acceptance ratio:
\begin{equation}
\frac{{Accep(\pi\pi\pi)}}{{Accep(\pi\mu e)}} = 1.86
 \pm 0.02(\rm stat.) \pm 0.03(\rm syst.)
\end{equation}
\begin{table}
\caption[Efficiencies used to calculate correction factor]   
{
Efficiencies used to calculate correction factor.
}
\begin{ruledtabular}
\begin{tabular}{lll}
 {\it{Effic.}}     &  {\it{Description}}  & {\it{Value}}          \\ \hline
$\epsilon_{\pi}$ & $\pi^+$ detection &$0.781 \pm 0.004$ \\[2mm]
$\epsilon_{\mu}$ &  $\mu^+$ detection  & $0.79 \pm 0.01$  \\[2mm]
$\epsilon_{e}$ &  $e^-$ detection  & $0.745 \pm 0.003$  \\[2mm]
$\epsilon_{M_{ee}}$ &  invariant mass cut
$M_{e^-\pi^{+}}>55$ MeV & \\
& and $M_{e^-\mu^{+}}>55$ MeV
& $0.919 \pm 0.007$  \\[2mm]
$\epsilon_{CL}$ & $e^-$ required on the left side & \\
& of the \v{C}erenkov counters  & $0.926 \pm 0.008$  \\[2mm]
$\epsilon_{CR}$ & $\pi^+$ required on the right
side & \\
& of the \v{C}erenkov counters  & $0.891 \pm 0.008$  \\[2mm] 
$\epsilon_{z}$ & reject $\pi^+$ from $\kpi2$ decay
&  0.865 $\pm$ 0.009 \\
\end{tabular}
\end{ruledtabular}
\label{tab:corr_fact}
\end{table}          

\subsection{Correction factors}
\label{sec:corf}
The correction factor $C$ in Eq. (\ref{eq:upper_limit_mc}) contains
efficiencies to account for the difference in reconstructing
$K_{\tau}$ and $K_{\pi\mu e}$ events. 
As described in  section \ref{sec:tau_total},
 most of the kinematic cuts for the $K_{\tau}$ sample selection
 were also applied
to select $K_{\pi\mu e}$ events and the corresponding efficiencies were
the same for both decay modes. The correction factor was calculated,
as a reciprocal of the the product of the selection
cut efficiencies specific
to the $K_{\pi\mu e}$ (see Table \ref{tab:corr_fact}), to be $C=3.32\pm 0.08$.
The bulk of $C$, 2.17, is the reciprocal of the
product of the $\pi^+$, $\mu^+$ and $e^-$  PID
efficiencies, while the remaining factor of 1.53 results from the specific
$K_{\pi\mu e}$ kinematic cuts.
\vspace*{0.5cm}

\section{Results and conclusions}
With the $B({\pi\pi\pi})=5.58\pm0.03\%$\cite{Pdg04}, and 
all the other factors in Eq. (\ref{eq:upper_limit_mc}) determined, the
upper limit on the $K_{\pi\mu e}$ branching ratio is set at:
\begin{equation}
B(\kpme) < 2.1 \times 10^{-11} \  (90\% \ \rm C.L.)
\end{equation}

         The details of the analysis of the final data set
          obtained by E865 are described in this paper. Here, we
          summarize the search for the decay $K^+\to\pi^+\mu^+ e^-$
          at BNL, which included a series of four measurements in which
         the following 90\% C.L. upper limits were obtained:
         $2.1\times10^{-10}$\cite{Lee}, $2.0\times10^{-10}$\cite{sd1995},
         $3.9\times10^{-11}$\cite{Hanh00,prl00}, and $2.1\times10^{-11}$
         (this paper). Since the last two measurements were not
         background  free, we analyzed the combined likelihood function
         constructed as the product of the
         likelihood functions for the separate measurements to establish
         a combined upper limit. For the final data set,
         this function is described above in Eq. \ref{eq:poisson_lik1}.
 The analogous
         function, based on the distributions shown in Fig. 4 of
         Ref. \cite{prl00} was used for the 1996 data. The only common
         parameter was a $K^+\rightarrow \pi^+\mu^+ e^-$ branching ratio $B$,
         introduced into the likelihood function via
\begin{displaymath}
         N_{sig}=N_K^{eff}B,
\end{displaymath}
         where $N_K^{eff}$ was 
	the effective (corrected for $K_{\pi\mu e}$ acceptance) number
  of kaon decays
         in a given experimental run.
  For the background and signal free
         measurements of Refs. \cite{Lee} (E777) and \cite{sd1995},
 the likelihood function (Eq. \ref{eq:poisson_lik1})
         is reduced to the simple form $\exp{(-N_K^{eff}B)}$.

        Using the Bayesian approach in the analysis of the combined
        likelihood  function, we have obtained
\begin{displaymath}
 B(K+\rightarrow \pi^+\mu^+e^-)<1.3\times10^{-11}\ \ (90\% \ \rm{C.L.})
\end{displaymath}

Although no evidence of new physics was found, 
 the parameter space of the existing extension theories that allow
LFNV was reduced.
Particularly, as discussed in Sec. \ref{sec:introduction},
 the mass, $M_H$, of the corresponding Extended Technicolor boson, with
strength
 equal to that of 
the weak interaction, is
bound by the limit obtained on the $K_{\pi\mu e}$ branching ratio
 to be greater then $\approx80$ TeV.

\begin{acknowledgments}

We gratefully acknowledge the contributions to the success of
this experiment by Dave Phillips,
the staff and management of the AGS at the Brookhaven National
Laboratory, and the technical staffs of the participating institutions.

Also we would like to acknowledge discussions
concerning this analysis with Dr. Andries van der Schaaf (Zurich)
and contributions to earlier stages of the experiment by Dr.
 Peter Robman (Zurich).

This work was supported in part by the U. S. Department of Energy,
the National Science Foundations of the USA, Russia and Switzerland, and
the Research Corporation.
\end{acknowledgments}


\end{document}